\documentclass[journal]{IEEEtran}

\usepackage{ifpdf}
\usepackage{amssymb}
\usepackage{graphicx}
\usepackage{epstopdf}
\usepackage[cmex10]{amsmath}
\usepackage{url}
\usepackage{algorithm}
\usepackage{algorithmic}
\usepackage{multirow}
\usepackage{subfigure}
\usepackage{xspace}

\newcommand{\nonBlocking}{{BAFEC}\xspace}
\newcommand{\multiclass}{{MBAFEC}\xspace}

\newcommand{\rateVec}{\hat{\Lambda}}
\newcommand{\codeVec}{\hat{N}}
\newcommand{\minCodeVec}{\hat{K}}
\newcommand{\usageVec}{\hat{U}}
\newcommand{\compVec}{\hat{\alpha}}

\newcommand{\Expect}{\mathbb{E}}

\newtheorem{theorem}{\textbf{Theorem}}
\newtheorem{corollary}{\textbf{Corollary}}

\begin{document}

\title{FAST CLOUD: Pushing the Envelope on Delay Performance of Cloud Storage with Coding}

\author{Guanfeng~Liang,~\IEEEmembership{Member,~IEEE,}
        and~Ula\c{s}~C.~Kozat,~\IEEEmembership{Senior Member,~IEEE}
\thanks{Accepted for publication in IEEE/ACM Transactions on Networking on October 30th, 2013.}
\thanks{G. Liang and U.C. Kozat are with DOCOMO Innovations Inc., Palo Alto, California USA. G. Liang is the contact author. E-mail: gliang@docomoinnovations.com}
}
\maketitle

\begin{abstract}
Our paper presents solutions that can significantly improve the delay performance of putting and retrieving data in and out of cloud storage. We first focus on measuring the delay performance of a very popular cloud storage service Amazon S3. We establish that there is significant randomness in service times for reading and writing small and medium size objects when assigned distinct keys. We further demonstrate that using erasure coding, parallel connections to storage cloud and limited chunking (i.e., dividing the object into a few smaller objects)  together pushes the envelope on service time distributions significantly (e.g., 76\%, 80\%, and 85\% reductions in mean, 90th, and 99th percentiles for 2 Mbyte files) at the expense of additional storage (e.g., 1.75$\times$). However, chunking and erasure coding increase the load and hence the queuing delays while reducing the supportable rate region in number of requests per second per node. Thus, in the second part of our paper we focus on analyzing the delay performance when chunking, FEC, and parallel connections are used together. Based on this analysis, we develop load adaptive algorithms that can pick the best code rate on a per request basis by using off-line computed queue backlog thresholds. The solutions work with homogeneous services with fixed object sizes, chunk sizes, operation type (e.g., read or write) as well as heterogeneous services with mixture of object sizes, chunk sizes, and operation types. We also present a simple greedy solution that opportunistically uses idle connections and picks the erasure coding rate accordingly on the fly. Both backlog and greedy solutions support the full rate region and provide best mean delay performance when compared to the best fixed coding rate policy. Our evaluations show that backlog based solutions achieve better delay performance at higher percentile values than the greedy solution.

\end{abstract}


\IEEEpeerreviewmaketitle

\section{Introduction}
\label{sec:intro}

Public clouds have been utilized by web services and Internet applications widespread. They provide high degree of availability, scalability, and data durability. 
Yet, there exists significant skew in network bound I/O performance necessitating solutions that provide robustness in a cost effective manner \cite{Garfinkel07anevaluation,Wang:2010}. In this paper, we focus on the cloud storage and present solutions that can provide much better delay performance for putting files into the cloud storage as well as for retrieving them back on demand. In particular, we base our analysis on Amazon S3 service as one of the most popular cloud storage services.

A typical cloud storage stores and retrieves objects via their unique keys. Each object is replicated several times within the cloud and sometimes also further protected by erasure codes to more efficiently use the storage capacity while attaining very high durability guarantees \cite{Huang12}. Storage provider also monitors the load on each storage node and employs dynamic load balancing to prevent hot storage nodes that might observe high loads or slow nodes that have excessively high response times. Although mainly used for repairing data in unavailable storage nodes, some cloud providers also access coded blocks in parallel to uncoded blocks when uncoded blocks are stored in slow nodes \cite{Huang12}. Despite all these mechanisms, still evaluations of large scale systems indicate that there is a high degree of randomness in delay performance \cite{Garfinkel07anevaluation}. Thus, the services that require better delay performance must deploy their own solutions such as sending multiple requests (in parallel or sequentially), chunking large objects into smaller ones and read/write each chunk in parallel, replicate the same object using multiple distinct keys, etc. 

To this end, we conducted our own measurements on Amazon S3 for various object sizes to model its delay distribution. Our measurement results confirm that the delay spread is significant even when object sizes are in the order of megabytes. Moreover, our study indicates that when the server accessing the storage cloud is not the bottleneck (in terms of CPU and network access speed),  we can substantially improve the distribution of read/write delays. To achieve these gains, one has to consider not only chunking and parallel access to each chunk, but also erasure coding. In fact without erasure coding, more chunking starts hurting the performance at lower percentile values. The gains when forward error correction (FEC) is employed are significant in the average delay performance and they are much better at higher percentile delays.  

\begin{figure}[!t]
\centering
\includegraphics[width=\columnwidth]{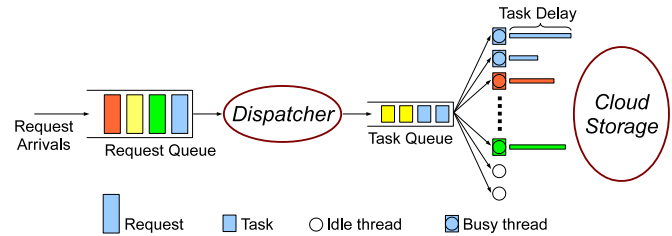}
\vspace{-20pt}
\caption{System Model}
\label{fig:system}
\vspace{-15pt}
\end{figure}

Nonetheless, server accessing the storage cloud has limited CPU and network access speed limiting the number of concurrent connections to the storage cloud without going into a processor sharing mode. With limited system capacity, one has to consider the load and its impact on queueing delays to quantify the total delay. Unfortunately, FEC and chunking create redundant load multiplying the arrival rate into the system. Unless mean service rate is improved at the same rate, the maximum rate at which end users can be served is reduced. Our observations over Amazon S3 indicate that indeed lower code rates reduce the supportable rate region inducing queue instability earlier than higher code rates. Thus, it is imperative to design a load adaptive strategy for changing FEC rates on the fly to keep total average delays at the minimum level while remaining in the achievable rate region of the uncoded system. 

To have meaningful solutions, one needs to analyze the queuing delay for the system. As one of the main contributions of the paper, we analyze the average delay performance of a system that incorporates chunking, FEC and multiple servers. This system model is much harder than an M/G/k queue, which itself have only crude approximations, as the service times of servers become interdependent due to the use of erasure coding. To make this point more clear, consider the case where an object is divided into two parts and a third part is generated by bit-wise XOR. If three servers are idle, then each part can be accessed in parallel. As soon as any two server complete their jobs, the third server can preempt its current job as erasure coding renders the completion of this job irrelevant. Except for a very recent work \cite{Longbocodeingincloud} that targets to solve a much simpler yet still hard case, to the best of our knowledge queuing analysis for such a system model is quite an uncharted area. Our analyses provide a good approximation for capacity and mean delay for homogeneous traffic with one operation type (e.g., all reads) and file size as well as for heterogeneous traffic with mixture of traffic types (e.g., both read and write requests with varying chunking and file sizes).

As another major contribution, we develop three load adaptive FEC schemes that change the coding rate on the fly. Using the analysis results, we can actually identify under what load regimes which fixed FEC strategy provides the best average delay performance leading to simple backlog threshold based adaptive algorithms. We present two schemes \nonBlocking (for single type of requests, i.e., homogeneous traffic) and \multiclass (for multiple types of requests, i.e., heterogeneous traffic) that adapt FEC rates based on the queue backlog. Via simulations using real service time traces from Amazon S3, we show that both schemes are able to beat the delay performance of any fixed FEC rate policy while achieving the rate region of the uncoded strategy. Since both \nonBlocking and \multiclass require a priori knowledge and put constraints on service time distribution of cloud storage to compute the optimal thresholds, we also propose a greedy strategy that opportunistically determines FEC rates based on the number of idle servers at the time of request arrivals.  Trace driven simulations demonstrate that the greedy strategy performs on a par with the queue backlog based strategies in terms of total mean delay. Nonetheless, the greedy method performs significantly worse in some cases at very high percentile values (e.g., at 99.9th percentile).

The remaining sections are organized as follows. In Section~\ref{sec:system}, we explain our system model in more details. In Section~\ref{sec:measurement}, we present our measurement results over Amazon EC2 and S3. In Section~\ref{sec:singletype}, we study the single-class scenario and develop a  FEC rate adaptive scheme \nonBlocking based on the analysis, and evaluate its performance through trace-driven simulations. In Section \ref{sec:theory}, we generalize the analysis to multi-class scenario and develop a multi-class FEC rate adaptive scheme \multiclass.  
In Section~\ref{sec:related}, we cover the related literature. Finally, we conclude the paper in Section~\ref{sec:conclusion}.

\section{Related Work}
\label{sec:related}

FEC in connection with multiple paths and/or multiple servers is a well investigated topic in the literature \cite{VickySharmaMPLOT,EminGabrielyanFEC,JohnByersAccessing,RSaadEvaluating}. However, there is very little attention devoted to the queueing delays. FEC in the context of network coding or coded scheduling has also been a popular topic from the perspectives of throughput (or network utility) maximization and throughput vs. service delay trade-offs \cite{Eryilmaz:2008:DTG:2263482.2273567,Yeownetworkcoding,Theodorosnetworkcoding, KozatScheduling}. Although some incorporate queuing delay analysis, the treatment is largely for broadcast wireless channels with quite different system characteristics and constraints.
FEC has also been extensively studied in the context of distributed storage from the points of high durability and availability while attaining high storage efficiency \cite{Dimakis:2010:NCD:1861840.1861868,Rodrigues_highavailability,Li:2010:TDR:1833515.1833884,ferner2012Allerton}.

Two papers \cite{Longbocodeingincloud,MDS-queue} concurrent to ours conducted theoretical study of cloud storage systems using FEC in a similar fashion as we do in this paper. Both papers rely on the assumption of exponential task delays, which hardly captures the reality. Therefore, some of their theoretical results are over optimistic and  cannot be applied in practice.
For example, authors of \cite{MDS-queue} proved that using larger code lengths always improves delay without reducing system capacity, contradicting with simulation results using real-world measurements presented in this paper.

Another set of works that is closely related to our work looks directly into the delay performance of storage clouds \cite{Garfinkel07anevaluation, stout}. The measurements results and interim conclusions in \cite{Garfinkel07anevaluation} on Amazon S3 motivated our work. The paper presents the throughput-delay tradeoffs in service times as object sizes vary. They establish the skewness and long tails. They recommend to cancel long pending jobs and send a fresh request instead. Although the suggestion would work well for long tails, this would not lead to much delay improvement below 99th percentile.  \cite{stout} on the other hand focuses more closely on the throughput-service delay tradeoff and devise a data batching scheme. Based on the observed congestion, authors increase or reduce the batching size. Thus, at high congestion, a larger batch size is used to improve the throughput while at low congestion a smaller batch size is adopted to reduce the delay.  The chunk size in our work is similar to the batch size considered in  \cite{stout} and it remains as a future work how to combine these complementary ideas.

\section{System Model}
\label{sec:system}

\subsection{Basic Architecture and Functionality}
The basic system architecture captures how web services today utilize public or private storage clouds.  The architecture consists of proxy servers in the front end and a key-value store (referred to as cloud storage) in the backend. 

Proxy servers have two main responsibilities: (1) Present a rich service layer  that operates on top of the raw cloud storage services/interfaces. (2) Optimize the user perceived performance. Client requests arrive at any of the proxy servers. When client wants to upload a file, proxy server divides the file into one or more chunks. Each chunk is stored as an individual object with a unique key in the key-value store. When the entire file is written successfully, the job is completed and a response is sent back to the client.  When client wants to download a file, proxy server checks which chunks need to be fetched from the storage cloud. Proxy generates read requests for these chunks and after receiving the complete set of chunks, the job is completed and the file is streamed back to the client. The solutions we present are deployed on the proxy server side transparent to the cloud storage.

Cloud storage has two main purposes: (1) Provide data storage with high durability and availability. (2) Provide on demand scaling of storage needs. Cloud storage does not interpret the objects it stores, but rather treats them as byte strings with a well-defined length. For high durability and availability, typical cloud systems replicate each object several times in different physical locations and may use FEC internally. From proxy servers' perspective, cloud storage is a black box whose internal techniques are unknown. Proxy servers only know the response times for each query (e.g., putting, getting, copying, deleting objects) it sends to the cloud storage.

\subsection{Adding FEC Support in Multi-threaded Proxies}
In our design, we employ maximum distance separable (MDS) codes \cite{lincostello}. Suppose a file is divided into $k$ equal size chunks (with padding). An (n,k) MDS code (e.g., Reed-Soloman codes) can expand these $k$ original chunks into $n\ge k$ coded chunks such that any $k$ chunks out of $n$ are sufficient to efficiently restore the $k$ original chunks (hence the file itself). 

MDS codes can help reducing the read delays as follows. Suppose proxy node have already segmented the requested file into $k$ chunks, expanded into $n_{max}$ chunks using an $(n_{max},k)$ MDS code, and written each chunk as a separate object using a unique key into the storage cloud.
When the file is to be read, proxy schedules  $n$ read tasks for distinct chunks using $n$ threads (not necessarily distinct ones) such that $k \le n \le n_{max}$.  Earliest $k$ successful responses from the storage cloud would then be sufficient to complete the read operation as $k$ chunks can be decoded to the original file chunks without requiring the remaining chunks (thus the read tasks for those chunks can be canceled). Notice that we implicitly assumed parallel independent task handling. If the tasks cannot be served in parallel or have strong correlation in their service latencies, FEC would impede the delay performance due to the extra load and processing overheads it generates.

Write operations are supported in a similar vein. Proxy can divide the file into $k$ chunks of equal size and encode them into $n$ coded chunks. The proxy then creates $n$ write tasks, one for each coded chunk. It schedules the tasks using $n$ threads. As soon as any $k$ of the $n$ uploading tasks complete, sufficient data has been stored on the cloud storage system.  
Thus, upon receiving $k$ successful responses from the storage cloud, the proxy sends a {\em speculative} success response to the client, without waiting for the remaining $n-k$ task to finish.
Such speculative execution is a commonly practiced optimization technique to reduce client perceived delay in many computer systems such as databases and replicated state machines \cite{zyzzyva}, etc.
Depending on the subsequent read profile on the same file, the proxy can (1) continue serving the remaining tasks till all $n$ tasks finish, or (2) change them to low priority jobs that will be served only when system utilization is low, or (3) cancel them preemptively. The proxy can even (4) run a demon program in the background that generates all $n_{max}$ coded chunks from the already uploaded chunks when the system is not busy.

We assume that subsequent read requests for an object that has been just written happens at time scales greater than the time necessary to commit all $n_{max}$ chunks. For the analysis in the later sections as well as for the proposed algorithms, this assumption is not a real limitation. In the case of performance results, the workloads that does not conform with this assumption would have a limited transient impact that will disappear when steady state performance is considered. Furthermore, in practice, such workloads are better handled by caching the most recently written objects in the proxies.

\subsection{Queueing Model with Multiple Threads and Coding}
Due to shared  resources, the level of parallelism achievable by using multiple threads is limited: the system can only support a finite number of simultaneously active threads without significantly degrading the performance of each individual active thread. Thus, we denote the maximum number of simultaneously active threads allowed in our system as $L$.
Under this constraint, we assume that the performance of each individual active thread is independent of the total number of active threads during the span of its life time.  

Accordingly, we model our proxy system by the queueing system shown in Fig.~\ref{fig:system}. There are two FIFO (first-in-first-out) queues in the system: one {\em request queue} that buffers all incoming requests that have not started yet, and one {\em task queue} that holds all waiting tasks of requests being served. $L$ threads are attached to the task queue. Whenever a thread becomes idle, it immediately starts serving the head-of-line (HoL) task in the task queue. The scheduler monitors the state of the queues and the threads, and decides what code rate should be used for each request in the request queue. The scheduler instructs the dispatcher to remove the HoL request from the request queue only if there is at least one idle thread. The dispatcher then creates the tasks for this request according to the code rate chosen by the scheduler, and injects them into the task queue. The idle threads immediately start serving (some of) the newly injected tasks. At the time when a request is completed, if some of its tasks are waiting, the waiting tasks are removed from the task queue. For a completed request,  if some of its tasks are still being served, they are canceled and the threads serving them become idle.

Depending on the criteria according to which the HoL request of the request queue should be admitted into the task queue, scheduling policies can be classified into the two categories below. Here, we assume that the scheduler has decided to serve the HoL request with an $(n,k)$ code.
\begin{itemize}
\item {\bf Blocking}: The HoL request is admitted into the task queue if and only if there are at least $n$ idle threads. 

\item {\bf Non-blocking}: The HoL request is admitted into the task queue if and only if there is at least 1 idle thread.
\end{itemize}

Blocking policies are not work conserving, thus waste system capacity for keeping  threads idle unnecessarily. However, it has a nice structure that facilitates tractable queue analysis and provides good approximation for non-blocking policies.

\subsection{Multiple Classes of Requests}

In general, applications receive requests for both reading and writing for files of various sizes. From our measurement results (next section), it can be seen  that the distributions of service times of tasks of different operation types and/or different chunk sizes differ significantly. Also, requests for different applications may have different delay targets (for example, video streaming has different delay requirements than uploading a document). As a result, it would be preferable to use different chunk sizes for different requests to accommodate different delay requirements. It is then natural to group requests that have the same operation type, similar file sizes and similar delay requirements into one {\em class} and consider a composition of $m\ge 1$ classes of requests. Details of modeling multiple classes of requests will be presented in Section \ref{ssec:model:delay}. The following discussion of this paper will concentrate on queue management and adaptation of the amount of redundant read/write operations, based on the assumption that classes are given and the corresponding file/chunk sizes are predetermined. Determining the choices of these parameters as functions of different delay requirements remains part of our future work.

\subsection{Definition of Delays}
Consider a time period $[0,T]$. We denote the  set of requests arrived during this period by $I=\{1,2,3,\cdots,r,\cdots,N_T\}$, where  $r$ denotes the $r$-th arrived request and $N_T$ is the total requests during the period. For each request $r$, denote $T_A^r$ as the time when it arrives into the system. Given that request $r$ is served with an $(n,k)$ code, we index the corresponding $n$ tasks from 1 to $n$, according to the time they start being served, and denote $T_S^{r,1}\le T_S^{r,2}\le \cdots \le T_S^{r,n}$ as their starting times. Also denote $T_F^{r,j}$ as the completion time of task $j$ of request $r$. Note that the tasks are only ordered by their starting times but not the completion times. So it is possible that $T_F^{r,j} > T_F^{r,l}$ even if $j<l$. 
The starting time of a request $r$, denoted as $T_S^r$, is defined as the time it gets admitted into the task queue, i.e., the starting time of its first task. So $T_S^r = T_S^{r,1}$. Its finish/completion time, denoted as $T_F^r$, should be the time when $k$ of its tasks have finished. Let $T_F^{r,1:n}\le T_F^{r,2:n}\le \cdots\le T_F^{r,n:n}$ be the sorted permutation of the finish times of request $r$'s tasks. Then $T_F^r = T_F^{r,k:n}$.

The {\em queueing delay} for request $r$ is the length of time that it spends waiting in the request queue, denoted by $D_q^r = T_S^r - T_A^r$. The {\em service delay} for request $r$ is the time it spends in the system getting served, denoted by $D_s^r = T_F^r - T_S^r$. 
We also denote the {\em task delay} for task $j$ of request $r$ by $D^{r,j} = T_F^{r,j} - T_S^{r,j}$ unless the task is canceled When the task is canceled (because $k$ other tasks for the same request have completed), $D^{r,j} = T_F^r - T_S^{r,j}$.

\section{Measurement Results and Delay Model}
\label{sec:measurement}

\subsection{Measurement Results}
To model the distributions of service times ($D^{r,j}$) of individual tasks, we run measurements over Amazon EC2 and S3. EC2 instance served as our proxy node in the system model.  We instantiated an extra large EC2 instance with high I/O capability in the same availability region as the S3 bucket that stores our objects. We run experiments within North California as well as Tokyo regions. We benchmarked single thread vs. multiple thread environments to measure the impact of thread contention. For the machine type we used we were able to run 16 threads in parallel with almost linear gain in system throughput and observed almost identical delay distribution as single-thread. This means that for up to 16 parallel threads the bottleneck is neither in the capacity of the EC2 instance nor in the network. We conducted experiments on different week days in March, April, June, and July 2012 with various packet sizes 128Byte, 1KB, 0.5MB, 1MB, 2MB, and 3MB using 16 threads in parallel while saturating each thread. Each experiment lasted around 24 hours. We alternated between different packet sizes to capture similar time of day characteristics across packet sizes. For the same reasons, we also alternated between write and read jobs by first creating a batch of write jobs using distinct keys, then creating a batch of read jobs for these distinct keys once all the writes are completed successfully. Due to lack of space, we only show a limited set of results although the cross-correlation properties and cumulative distribution functions exhibit similar properties. We briefly present a representative subset of our main findings.  

\begin{figure}[!t]
\centering
\includegraphics[width=0.7\columnwidth]{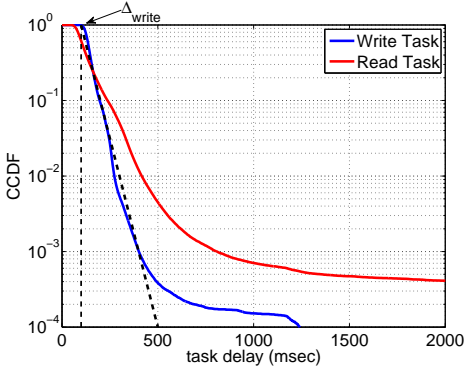}
\vspace{-10pt}
\caption{CCDF of $D^{r,j}$ for read \& write tasks for 1MB chunk.}
\label{fig:ccdf}
\vspace{-10pt}
\end{figure}

Fig.~\ref{fig:ccdf} plots the complementary cumulative distribution function (CCDF) of $D^{r,j}$ for read and write  tasks of 1MB chunks. Note that we only measure the time spent in any thread and there are no queuing delays. Read tasks for small to medium object sizes experience lower mean and median delays than the write tasks, yet at higher percentile delays (in this plot beyond 80th percentile) reads observe higher delays. Although not shown, as object size gets smaller the crossover point moves towards higher delay percentiles.

We also observe negligible correlation between the service times of subsequent tasks: the Pearson's correlation coefficient between the $t$-th and $(t+\tau)$-th tasks is always $\le 0.05$ for all $\tau\neq 0$.
This observation is critical as FEC techniques would be too costly and with little benefit if there were a strong correlation. The observation holds for all the packet sizes we experimented with as well as for the write tasks. Based on these results, for further analysis, we will treat task service times as independent and identically distributed (i.i.d.).

\begin{figure}[!t]
\centering
\includegraphics[width=0.7\columnwidth]{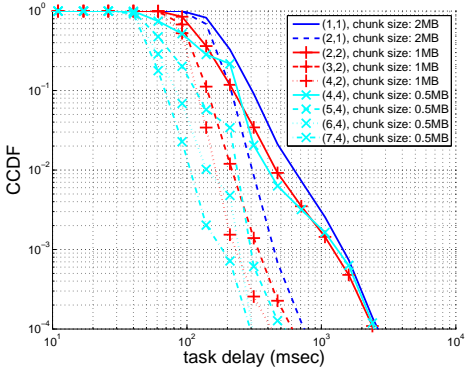}
\vspace{-10pt}
\caption{CCDF of service times for reading 2MB file} 
\label{fig:servicetimes}
\vspace{-10pt}
\end{figure}

To show the impact of using different codes on the service times (i.e., $D_s^r$ as opposed to $D^{r,j}$), we  plot the case for 2MB files with codes ranging from $(1,1)$ to $(7,4)$ in Fig.~\ref{fig:servicetimes}. Codes $(1,1), (2,2), (4,4)$ do not employ FEC, but instead use different chunk sizes. $(2,1)$ code provides 23\%, 32\%, and 56\% reduction in mean, 90th percentile, and 99th percentile delays over $(1,1)$ using 2$\times$ more storage. 
Using smaller chunk sizes with FEC improves delays at the same or less storage cost. E.g., (3,2) code provides 
50\%, 55\%, and 69\% reductions in mean, 90th percentile, and 99th percentile delays over $(1,1)$ using 1.5$\times$ storage, (5,4) code gives more than 60\% reductions in the same percentiles using only 1.25$\times$ storage, and (7,4) code improves delays by 76\%, 80\%, and 85\% at the expense of 1.75$\times$ storage.
Using smaller chunk sizes without FEC improves mean delay performance, but at higher percentiles the benefits deteriorate. This is expected as uncoded chunking requires completion of all tasks and small chunk sizes also have a long tail. The chances of catching the tail increases as the number of chunks increases.  FEC greatly mitigates this all or nothing behavior. 
The gains in service delay $D_s$ is only half of the story as chunking and FEC both adversely affect the achievable rate region as examined in later sections.

\subsection{Model of Task Delays}
\label{ssec:model:delay}
From Fig.~\ref{fig:ccdf}, it can be observed that for both read and write tasks, despite the delay floors observed at very low percentiles,  up to 99th percentile and even beyond that, 
the CCDF is roughly a constant term (which probably results from unavoidable overheads in any storage system such as networking delay, protocol-processing, lock acquisitions, transaction log commits, etc.) plus a linearly decaying term in log scale (which is a signature for distributions having an exponential tail). 
So we decide to model the task delays as i.i.d. random variables in the form of $\Delta + D_{exp}$, where $\Delta$ is a non-negative constant (corresponding to the constant term in CCDF), and $D_{exp}$ is an exponentially distributed random variable with some mean $1/\mu$ (corresponding to the linear term in CCDF). 
For mean delay analysis, our simulations later will show that this approximation works reasonably well.

We assume there are $m\ge 1$ classes of requests. Requests of each class have identical file size  and all are divided into chunks of identical size. Under this assumption, service times of all chunks of the same class follow the same distribution and each class $i$ can be characterized by a three-tuple $(k_i,\Delta_i,\mu_i)$, where $\Delta_i$ and $\mu_i$ specifies the delay distribution of class-$i$ chunks.
Throughout this paper, we assume $k_i$'s (and accordingly chunk sizes) are determined a priori and $(\Delta_i,\mu_i)$ are given. Our focus will be on the adaptation/choice of $n_i$'s.

\newcommand{\appxC}{\tilde{C}}
\newcommand{\appxD}{\tilde{D}}

\section{Single-Class (Homogeneous) Arrivals}
\label{sec:singletype}
In this section, we study the scenario when there is only one class of request, i.e., $m=1$. Since there is only one class, we will drop the subscript $i$ within this section.

We first investigate the delay and throughput tradeoff with fixed FEC, i.e., a fixed $(n,k)$ code is used for all requests, for both blocking and non-blocking schemes. Due to the interdependent nature of task delays while employing FEC, the queueing model for these policies is much more complicated than M/G/k queue, which itself has only crude approximations for delays. We are not able to provide exact analysis at this time. However, we develop reasonable approximations for both capacity and delay of these policies. Based on these approximation results, we develop a backlog-based adaptive FEC scheduler \nonBlocking, which achieves the best delay performance against fixed FEC schemes for all supportable arrival rates.

\begin{figure}[t]
\centering
\includegraphics[width=\columnwidth]{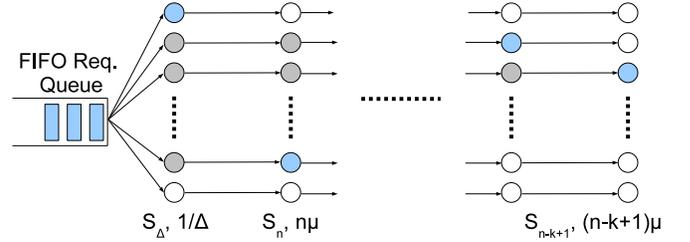}
\vspace{-20pt}
\caption{Multi-phase queueing model. A blue server indicates a request is being served at the corresponding phase and the corresponding pipe is occupied. The numbers at the bottom of each phase are the number of busy servers and the service rate of each server of that phase.}
\label{fig:blocking}
\vspace{-10pt}
\end{figure}

\subsection{Queueing Model for Blocking Policies with Fixed FEC}
Given our assumption that task delay is in the form of $\Delta + D_{exp}$, it can be considered that after started being served by a thread, a task experiences two phases of services: first a fixed-time service for $\Delta$, then followed by an exponential-time service with mean $1/\mu$. Recall that in blocking policies, all tasks of a request $i$ start at the same time, i.e., $T_S^{r} = T_S^{r,1}=\cdots T_S^{r,n}$. 
Then the service received by each request can be modeled in $k+1$ phases. The first is a fixed-delay phase of length $\Delta$, while all $n$ tasks are in their fixed-time service phase. The second is an exponential phase with mean $1/n\mu$, while all $n$ tasks are receiving exponential-time service and one task finishes by the end; Similarly, the third is an exponential phase with mean $1/(n-1)\mu$, while the remaining $n-1$ tasks are receiving exponential-time service and one more task finishes by the end; $\cdots$; the ($k+1$)-th is an exponential phase with mean $1/(n-k+1)\mu$, while the last $n-k+1$ tasks receiving exponential-time service and the $k$-th task finishes by the end (hence the whole request finishes and the remaining tasks are canceled). We will say a request is in phase $\Delta$ or phase $j$ ($n-k+1\le j \le n$) depending on the number of its remaining tasks and the phase these tasks are in. 
Now we can model a blocking policy with the queueing system depicted in Fig.~\ref{fig:blocking}. 
The FIFO request queue is followed by a set of parallel pipes of servers. Each pipe consists $k+1$ servers that represents the $k+1$ phases a request experiences during service: the first server has fixed service time $\Delta$ consuming $n$ active threads, the second has exponential service time with mean $1/n\mu$ consuming $n$ active threads, ..., the $(k+1)$-th has exponential service time with mean $1/(n-k+1)\mu$ consuming $n-k+1$ active threads. At any time, a pipe can be ``occupied'' by at most one request, i.e., at most one of its servers can be active. There are $\lceil L/(n-k+1)\rceil$ pipes in total\footnote{$\lceil L/(n-k+1)\rceil$ is the maximum number of requests that can be served simultaneously by a blocking policy since every request being served consumes at least $n-k+1$ active threads.} so that there will always be at least one unoccupied pipe as long as there are $\ge n$ idle threads, no matter which requests these threads were serving previously. 
Denote $S_{\Delta}(t)$ and $S_j(t)$  ($n-k+1\le j \le n$) as the number of requests being served in the corresponding phases at time $t$. Then the number of active threads at $t$ is $nS_{\Delta}(t) + \sum_{j=n-k+1}^n jS_j(t)$.
According to the definition of a blocking policy, all unoccupied pipes will be {\em blocked} for admission if $<n$ threads are idle. Whenever  
$$
nS_{\Delta}(t) + \sum_{j=n-k+1}^n jS_j(t) \le L-n,
$$
the unoccupied threads will be unblocked and the HoL request in the request queue will be admitted into one of them.

\subsection{Capacity of Blocking Policies}
\label{ssec:capBlocking}
Let $\overline{S}_{\Delta}$ and $\overline{S}_j$ denote the time average of  $S_{\Delta}$ and $S_j$. 
Assuming the queueing system is stabilized at arrival rate $\lambda$ and noticing that arrival rate to each phase equals to $\lambda$ when the system is stable, we have the following flow-balance equations from Little's law:
\begin{align*}
\overline{S}_{\Delta}  = \lambda\Delta
\text{~~~and~~~}
\overline{S}_j =  \frac{\lambda}{j\mu},~\forall n-k+1\le j \le n .
\end{align*}
As a result, the expected number of simultaneously active threads at arrival rate $\lambda$ is
\begin{equation*}
n\overline{S}_{\Delta} + \sum_{j=n-k+1}^n j\overline{S}_j = \lambda(n\Delta+k/\mu).
\end{equation*}
Since there are at most $L$ parallel active threads allowed, we have the following constraint on supportable arrival rates:
\begin{equation}
\lambda(n\Delta+k/\mu) \le L.
\label{eq:blocking:thread-bound}
\end{equation}

For the study of  capacity, 
defined as the maximum supportable arrival rate $\lambda$ of the system, it suffices to consider the case when the system is always backlogged. When always-backlogged, whenever there are at least $n$ idle threads, the HoL request will be admitted into one pipe. So the number of active threads is kept $\ge L-n+1$. Then we have the following upper and lower bounds on $C_b(n,k)$, the capacity of blocking policies using a fixed $(n,k)$ code:
\begin{equation}
\frac{L-n+1}{n\Delta + k/\mu }\le C_b(n,k) \le \frac{L}{n\Delta + k/\mu}.
\label{eq:blocking:cap-bound}
\end{equation}
While more accurate approximation is possible, we use the mean of  the two bounds as our estimation for $C_b$:
\begin{equation*}
\appxC_b(n,k) 
=\frac{L - (n-1)/2}{n\Delta + k/\mu}.
\end{equation*}

From the above discussion, we can see the capacity with fixed FEC is roughly proportional to the inverse of $$u(n) = n\Delta + k/\mu.$$ In fact, from our delay model, one can easily verify that $\Expect[\sum_{j=1}^n D^{r,j}] = n\Delta + \sum_{j=n-k+1}^n \frac{j}{j\mu} =u(n) $ (note that the slowest $n-k$ threads are canceled by the time $k$ threads finish). In other words, $u(n)$ is the expected sum of the amount of time used by $n$ threads in serving one request. For this reason, we call $u(n)$ the expected per-request system usage for using $(n,k)$ FEC code, or {\em usage} for short. The first term is linear in $n$ and represents the constant per-thread cost $\Delta$ we pay for having more parallelism. As we can see from Eq.\ref{eq:blocking:cap-bound} (especially upper bound), if $\Delta$ is large compared with $1/\mu$, the capacity is significantly reduced when a low rate FEC code (large $n$) is used and the queueing delay will quickly explode even at low arrival rate with respect to the capacity with no coding $C_b(k,k)$. We are going to investigate the delay issue in more detail in the rest of this section.

\subsection{Delays of Blocking Policies}
According to our model for task delay, the expected service delay of a blocking policy is $D_s(n,k) = \Delta +\sum_{j=n-k+1}^n \frac{1}{j\mu}$.

For queueing delay, we approximate the request queue and dispatcher by a virtual single-server queue. The virtual server's service time for request $r$ is determined by $T_S^{r+1} - T_S^r$, i.e., the inter-starting time of the requests $r$ and $r+1$ in our original system. So from the request queue's point of view, the virtual server behaves exactly as the dispatcher, and the virtual queue has the same queueing delay as our original system.

In general, the service times of different requests in the virtual system are not necessarily independent. In fact, the service time also depends on the arrival process. 
So the exact analysis of the queueing delay is very complicated. 
We notice that at low utilization, a request will most likely find enough idle threads to start being served immediately upon arrival, as if arriving at an empty M/G/1 queue. On the other hand, when utilization is higher, the system is mostly backlogged and the inter-starting times are weakly correlated because they are determined by the rate at which busy threads become idle. Based on these observations, we use an M/G/1 queue to approximate the behavior of the request queue, wherein the service times follow an Erlang distribution with parameters $n$ and mean $1/\appxC_b$.

To understand the choice of Erlang distribution, consider the case when $\Delta=0$, $n=k$.
Suppose the system is backlogged and all $L$ threads are busy immediately after $T_S^r$. Then $T_S^{r+1}$ is the time when the earliest $n$ out of $L$ threads become idle. Since $\Delta=0$, all task delays are exponential. Then the inter-starting time is the sum of $n$ exponential random variables, whose means are $1/(L)\mu,\cdots,1/(L-n+1)\mu$. This is very similar to an Erlang distribution with parameter $n$ and mean $\sum_{j=0}^{n-1}1/(L-j)\mu$, which is the sum of $n$ i.i.d. exponential random variables with mean $\frac{\sum_{j=0}^{n-1}1/(L-j)\mu}{n}$. When $L/n$ is sufficiently large, the inter-starting time distribution converges to the Erlang distribution.
When $\Delta> 0$ and $n>k$, this approximation can be quite crude. But we believe it is good enough as a guideline for policy design. Moreover, it also provides a simple closed-form approximation of the queueing delay, which is used in design of our adaptive FEC scheduler. 
Given an Erlang random variable  $X$ with parameter $n$ and mean $1/\appxC_b$, its second moment $\Expect[X^2] = (1+1/n)/\appxC_b^2$. Then queueing delay of the aforementioned M/G/1 queue (using the Pollaczek-Khinchin formula) is
\begin{align*}
\tilde{D}_q^b(n,k,\lambda) = \frac{\lambda \mathbb{E}[X^2]}{2(1-\lambda \mathbb{E}[X])}
 =\frac{\lambda(n+1)}{2n\appxC_b(n,k)(\appxC_b(n,k)-\lambda)}.
\label{eq:blocking:apprxDelay}
\end{align*}

\begin{table}[t]
\centering
\begin{tabular}{|c|c|c|c|c|}
\hline
\multirow{2}{*}{$\frac{\Delta}{\Delta + 1/\mu}$} & \multicolumn{2}{c|}{ Blocking, $L = 16$}  & \multicolumn{2}{c|}{ Blocking, $L = 64$}\\
\cline{2-5}
& $n=3$ & $n=6$ & $n=3$  & $n=6$\\
\hline
0.2& 1.0 -- 11.4 & 2.0 -- 20.6 & 0.3 -- 6.1 & 0.5 -- 8.6\\
\hline
0.4& 1.0 -- 13.4 & 0.8 -- 7.9 & 0.3 -- 7.8 & 0.5 -- 9.5\\
\hline
0.6& 1.2 -- 15.5 & 1.9 -- 63.2 & 0.3 -- 9.6 & 0.6 -- 11.7\\
\hline
0.8& 1.3 -- 18.0 & 1.0 -- 339.3 & 0.4 -- 11.6 & 0.6 -- 10.5\\
\hline \hline
\multirow{2}{*}{$\frac{\Delta}{\Delta + 1/\mu}$} & \multicolumn{2}{c|}{ Non-blocking, $L = 16$}  & \multicolumn{2}{c|}{Non-blocking,  $L = 64$}\\
\cline{2-5}
& $n=3$ & $n=6$ & $n=3$  & $n=6$\\
\hline
0.2& 0.9 -- 11.1 & 1.8 -- 8.5 & 0.3 -- 7.4 & 0.5 -- 8.4\\
\hline
0.4& 1.0 -- 13.1 & 1.9 -- 11.0 & 0.3 -- 8.2 & 0.5 -- 9.5\\
\hline
0.6& 1.1 -- 15.0 & 2.0 -- 16.8 & 0.3 -- 9.0 & 0.5 -- 10.4\\
\hline
0.8& 1.2 -- 16.4 & 2.1 -- 29.2 & 0.3 -- 10.0 & 0.6 -- 11.0\\
\hline
\end{tabular}
\caption{Range of errors: $|D_{sim}-\appxD|/\appxD \times 100\%$}
\label{tab:error}
\vspace{-25pt}
\end{table}

\subsection{Approximations for Non-Blocking Policies with Fixed FEC}
The only difference between blocking and non-blocking policies is that non-blocking policy starts a task whenever a thread becomes available, while blocking policy waits until $n$ threads become available. 
This difference is subtle yet it makes non-blocking policies much harder than blocking policies for exact analysis. In this section, we derive approximations of the capacity and delays of non-blocking policies. 

Notice that, when there are $L$ busy threads, the rate at which any single thread becomes available is in the order of $O(L/(\Delta + 1/\mu))$, which is much higher than the rate at which one particular busy thread becomes idle when $L$ is large. As a result, it is highly likely that, in a non-blocking policy, all tasks of a request will get started before any one of them finishes, and the gap between the first and last starting times of tasks are much smaller than the individual task delay. As a result, for large $L$,  a non-blocking policy behaves very similarly to a blocking policy that uses the same FEC code. Hence, the capacity of a non-blocking policy can be approximated by the capacity of a blocking one. Further notice that, when always backlogged, non-blocking policies always keep all $L$ threads busy. So we approximate the capacity of non-blocking policy with the upper bound for blocking:
\begin{equation}
\appxC_{nb}(n,k) = L/(n\Delta + k/\mu).
\label{eq:nonblocking:cap}
\end{equation}
Then we again use Pollaczek-Khinchin formula to estimate the queueing delay of non-blocking policy $\appxD_q^{nb}$, by replacing $\appxC_b$ with $\appxC_{nb}$ in the previous formulation for $\appxD_q^b$, and use $D_s=\Delta +\sum_{j=n-k+1}^n \frac{1}{j\mu}$ as an approximation of the service delay. By doing this, $\Expect[X] = u(n)/L$ for non-blocking. 

We compare the approximated delay of blocking and non-blocking policies $\appxD^b = D_s + \appxD^b_q$ and $\appxD^{nb} = D_s + \appxD^{nb}_q$ against the average delay from simulations using  task delays in the form of $\Delta+D_{exp}$ (denoted as $D_{sim}^b$ and $D_{sim}^{nb}$). 
Table \ref{tab:error} shows the range of estimation errors for $k=3$, $n=3,6$ and $L=16,64$, while arrival rate varies from $0.1\appxC_x$ to $0.9\appxC_x$ ($x=b$ or $nb$). 
For each setting, the lower end of estimation error is observed at low to medium arrival rates while larger error is observed for arrival rates near the estimated capacity. This is mainly due to the high sensitivity of $\appxD$ in  $\appxC$ when $\lambda \rightarrow \appxC$  (because of the  $\appxC - \lambda$ term in the denominator), so even a small discrepancy between $\appxC$ and the actual capacity will be significantly magnified in delay at arrival rate close to capacity. 
As we can see, the approximations are quite reasonable, except for the cases when $L=16$, $n=6$ and $\frac{\Delta}{\Delta + 1/\mu}=0.6,~0.8$. This is because Erlang distribution is not a good approximation for the inter-starting times when $\Delta$ and $n$ are large compared to $1/\mu$ and $L$, respectively.
We also observe that approximation for non-blocking policy is generally better than the one for blocking policy. 
This is because the Erlang distribution is a much better approximation for the inter-starting times of non-blocking schemes since the number of busy threads remains fixed (equals to $L$) when the system is backlogged.

\begin{figure}[t]
\centering
\includegraphics[width=0.7\columnwidth]{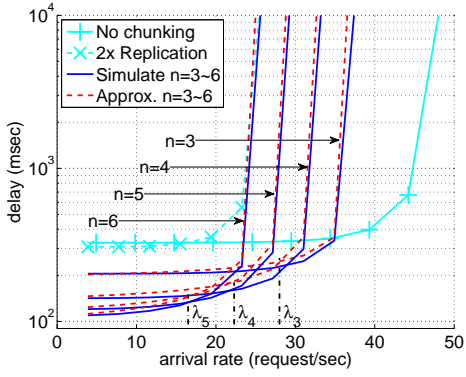}
\vspace{-10pt}
\caption{Estimation vs.  trace-driven simulation}
\label{fig:estimate_VS_trace}
\vspace{-10pt}
\end{figure}

We further compare the approximation against trace-driven simulations.  Fig.~\ref{fig:estimate_VS_trace} plots $\appxD^{nb}$ and the average delay from  simulations for reading 3MB files with fixed FEC schemes with $k=3$, $n=3,4,5,6$ and $L=16$, using traces for read operations we collected in March 2012 and chunk sizes of 1MB. For computation of $\appxC_{nb}$, we first filter out the worst 0.1\% task delays in the trace, then we set $1/\mu$ and $\Delta + 1/\mu$ as the standard deviation and the mean of the remaining task delays, respectively. 
We emphasize that although we use the filtered task delays to obtain estimations of $\Delta$ and $1/\mu$, all unfiltered task delays are used in the simulations. 
As we can see, our approximation matches the simulation results very well, which justifies our $\Delta +D_{exp}$ model for task delays. The simulation results also suggest that the capacity of non-blocking policies with fixed FEC is a decreasing function of $n$, which is consistent with our approximation of $\appxC^{nb}$ from Eq.\ref{eq:nonblocking:cap}. We also plot the delay for the simple no chunking solution (using $(1,1)$ code), as well as simple $2\times$ replication solution (using $(2,1)$ code) using traces for  chunk size 3MB collected in the same time period. Despite providing a larger capacity, the simple no chunking solution has very bad delay performance. Even for very low arrival rates, the delay is over 300 ms, while just chunking without FEC ($n=3$) improves the delay to about 200 ms with zero storage overhead and using a $(4,3)$ code with 1/3 storage overhead improves the delay to less than 150 ms. Moreover, the simple replication of unchunked objects not only fails in improving the delay but also significantly reduces the capacity.
This is because read/write operations for large object has a large constant overhead $\Delta$ and a relatively small delay spread $1/\mu$ according to our measurement results. So there is not enough diversity to gain from parallelism.
This again justifies our motivation for using chunking with FEC for delay sensitive applications. With the same amount of storage overhead, using a $(6,3)$ code delivers roughly $3\times$ improvement in delay.

\subsection{\nonBlocking -- Backlog-based Adaptive FEC Scheduler }
In this section, we present \nonBlocking -- a backlog-based adaptive FEC scheme that achieves the best delay achievable by any fixed FEC scheme with $k\le n\le n_{max}$, i.e.,
\begin{equation*}
\min_{k\le n\le n_{max}}\left( D_s(n,k) + \tilde{D}_q(n,k,\lambda)\right),
\end{equation*}
for all supportable arrival rates. The following discussion applies to both blocking and non-blocking policies, so we drop the superscript in the delay terms.  
Assuming $k$ is fixed, our estimation of the expected total delay  is a function of  $n$ and $\lambda$: $\tilde{D}(n,\lambda) = D_s(n) + \tilde{D}_q(n,\lambda)$. For every $n = k,\cdots, n_{max}-1$, we compute the solution $\lambda_n$ such that
\begin{align}
\tilde{D}(n,\lambda_n) = \tilde{D}(n+1,\lambda_n).
\label{eq:threshold}
\end{align}
In the example of Fig.~\ref{fig:estimate_VS_trace} we show $\lambda_3$ to $\lambda_5$, which are the intersection of the red dashed lines.
According to our previous analysis, it only requires solving a quadratic equation of $\lambda$ and only the smaller solution is meaningful. Due to limitation of space, we would not include the details.   
$\lambda_n$ is the crossover point for the delay performance of a $(n,k)$ code and a $(n+1,k)$ code: if $\lambda < \lambda_n$, then a $(n+1,k)$ code gives smaller total delay than a $(n,k)$ code does; and if $\lambda > \lambda_n$, a $(n,k)$ code will give smaller total delay.  Using Little's law, we compute the corresponding crossover backlog size $Q_n = \lambda_n \tilde{D}_q(n,\lambda_n)$. It is easy to show that $Q_n$ is a decreasing function of $n$, then we can use $\{Q_n\}$'s as thresholds to adapt the FEC code length based on the backlog size. The adaptive scheme is described formally as follows:

\vspace{5pt}
\hrule
\vspace{2pt}
\noindent \textit{\textbf{\nonBlocking (Backlog-based Adaptive FEC)}}
\hrule
\vspace{2pt}

\noindent Do the following for every request $r$

\begin{algorithmic} [1]

\STATE $\overline{Q} \leftarrow $ backlog size upon arrival of request $r$.
\STATE Find $n$ such that $\overline{Q}\in [Q_{n}, Q_{n-1})$, or $\overline{Q}\in [Q_{n},\infty)$ for $n=k$, or $\overline{Q}\in [0,Q_{n})$ for $n=n_{max}$.
\STATE Serve request $r$ with an $(n,k)$ code when it becomes HoL.
\end{algorithmic}
\hrule

~

Under our delay model, for a given $k$, $n=k$ provides the largest capacity region. Then \nonBlocking is in fact throughput-optimal, i.e., it supports any arrival process supportable by some $(n,k)$ code when $k\le n\le n_{max}$, because it always sets $n=k$ if the queue exceeds a finite threshold $Q_n$.   

~

\begin{figure}[t]
\centering
\includegraphics[width=0.7\columnwidth]{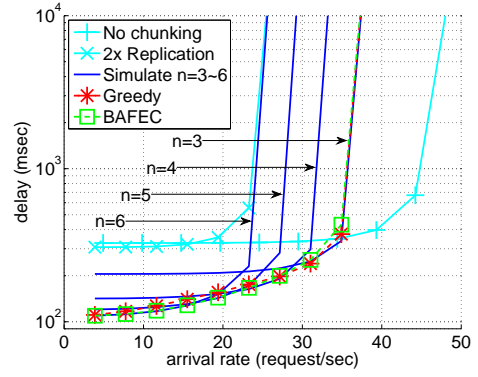}
\vspace{-10pt}
\caption{Average delay, fixed FEC vs. greedy vs. \nonBlocking}
\label{fig:delay:mean}
\end{figure}

\begin{figure}[t]
\centering
\includegraphics[width=0.7\columnwidth]{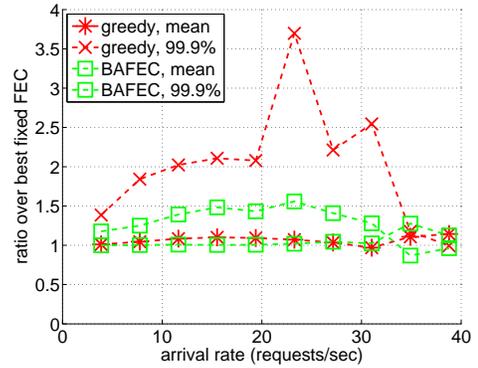}
\vspace{-10pt}
\caption{Greedy vs. \nonBlocking, normalized delays}
\label{fig:delay:ratio}
\vspace{-10pt}
\end{figure}

\subsection{Performance Evaluation}
We conduct trace-driven simulations for performance evaluation. Due to lack of space, we only show results for non-blocking versions, with $k=3$, $n_{max} = 6$ and $L=16$, using traces we collected in March 2012 and chunk size 1MB. Results for other settings of parameters and blocking versions are similar.
We also develop a simple {\em Greedy} heuristic scheme. Unlike \nonBlocking, Greedy does not require any knowledge of the distribution of task delays, yet it achieves competitive mean delay performance. In Greedy, the code used to serve request $r$ is determined by the number of idle threads upon its arrival: if there are $l \ge k$ idle threads, use a $(\min(l,n_{max}),k)$ code; otherwise use a $(k,k)$ code, i.e., no coding.

Fig.~\ref{fig:delay:mean} plots the average delays of fixed FEC schemes with $n=3,4,5,6$, as well as the delays of  Greedy  and \nonBlocking. As we can see, Greedy and \nonBlocking have almost identical performance in terms of average delay. Both adaptive schemes succeed in (roughly) achieving our goal of obtaining the lower envelop of the delay performance of the set of fixed FEC schemes. At lower utilization, they deliver over $3\times$ lower delay compared no chunking and simple $2\times$ replication ($n=1,2;k=1$), and $2\times$ lower delay compared to naive chunking ($n=k=3$).

We plot the average and 99.9\% delays of Greedy and \nonBlocking, normalized by the best delays obtained from fixed FEC schemes, in Fig.~\ref{fig:delay:ratio}. At very low and high arrival rates, these two adaptive schemes perform almost the same as the optimal fixed FEC scheme. This is because (1) with low arrival rates, there are no backlog most of the time and both schemes behave like a fixed FEC scheme with $n=n_{max}$; and (2) with high arrival rates, the system is always backlogged and both schemes behave like a fixed FEC scheme with $n=k$. In the intermediate region, \nonBlocking still traces the best performance of fixed FEC schemes very well, as it is almost identical to the best fixed FEC scheme in mean delay, and it stays within  $1.5\times$ of the optimal 99.9\% delay. On the other hand, while Greedy also achieves almost optimal mean delay performance, it performs much worse for high percentile delays. For low to medium arrival rates, Greedy is consistently above $2\times$ and can even go beyond $3.5\times$ of the optimal 99.9\% delays.

\newcommand{\smallwidth}{0.26\textwidth}

\section{Multiple-Class (Heterogeneous) Arrivals}
\label{sec:theory}
In this section, the scenario with multiple classes of requests ($m>1$) is studied. As the multi-class problem is even more complicated than the single-class one, we again based our analysis on the approximations of queueing and service delays. Our analysis shows that the delay-optimal combination of code lengths ($n_i$'s) has a well-defined structure that is helpful for designing practical rate adaptation schemes:
\begin{itemize}
\item There is an one-to-one mapping between the optimal code lengths and the corresponding expected total queue length (all classes combined), irrespective of the arrival rates;
\item The optimal code length of any class is a decreasing function of the expected total queue length. 
\end{itemize}
These analysis results suggest that (1) expected queue length is a good indicator of the optimal code lengths and (2) adaptation of each class can be done separately.
Based on these insights, we
develop a Multi-class Backlog-based Adaptive FEC (\multiclass) scheme. In \multiclass, each class $i$ is associated with a set of thresholds computed using Eq.\ref{eq:threshold} as in \nonBlocking, assuming  the single-class scenario with only class-$i$ requests;  and code adaptation within each class is performed in the same way as in \nonBlocking. 

\subsection{Fixed FEC Code Analysis}
\label{ssec:multi:fixed}
We assume that arrivals of each class $i$ follows a Poisson process with rate $\lambda_i>0$, independent of other classes. So the combined arrivals consist a Poisson process at rate $\lambda = \sum_{i=1}^m \lambda_i$. The following notations and terminologies will be used for the subsequent discussion.

\begin{itemize}
\item The (column) {\em rate vector} $\rateVec = [\lambda_1;\cdots;\lambda_m]$ and the {\em composition vector} $\compVec = [\alpha_1;\cdots;\alpha_m] = \rateVec /\lambda$. Note that $0< \alpha_i = \lambda_i/\lambda \le 1$ and $\sum_{i=1}^m \alpha_i = 1$. 

\item 
The {\em code vector} $\codeVec=[n_1;\cdots;n_m]$, given that $n_i$ is the code length chosen for class-$i$ requests.

\item 
The {\em usage vector} $\usageVec(\codeVec)=[u_1(n_1);\cdots;u_m(n_m)]$, where $u_i(n_i) = n_i \Delta_i + k_i / \mu_i$ is the per-request usage of class-$i$ requests. When it is clear from context, we will omit the function inputs ($\codeVec$ and $n_i$). 
\end{itemize}

We can easily generalize the multi-phase queueing model introduced in the previous section (Fig.~\ref{fig:blocking}) to incorporate multiple classes of requests. There is still one FIFO request queue, but instead of only one type of pipes, we construct a set of pipes for every class, with the number of servers in each pipe and their service rates specified by the delay parameters of the class. A class-$i$ request is admitted into a pipe for class $i$ only if there are  $\ge n_i$ idle threads. 
According to Little's law, we obtain flow-balance equations in the same vein as Section \ref{ssec:capBlocking}. Similar to Eq.\ref{eq:blocking:thread-bound}, for a given code vector $\codeVec$, a supportable rate vector $\rateVec$ must satisfy
\begin{equation*}
\sum_{i=1}^m \lambda_i(n_i \Delta_i + k_i/ \mu_i) 
= \rateVec^T \usageVec(\codeVec) 
= \lambda \compVec^T \usageVec(\codeVec) \le L
\label{eq:multiclass-rate-region}
\end{equation*}
for system stability. Starting from this point, we only consider  non-blocking (work conserving) policies. Similar to the single-class scenario, where with Eq.\ref{eq:nonblocking:cap} the capacity region is approximated by $\{\lambda:\lambda(n\Delta + k/\mu) \le L\}$, we approximate the multi-class capacity region {\em with respect to $\codeVec$} by the convex set
\begin{equation*}
C(\codeVec) = \{\rateVec: \rateVec^T\usageVec(\codeVec) \le L\},
\label{eq:multiclass-capacity}
\end{equation*}
and the capacity for a given composition of requests $\compVec$ is
\begin{equation*}
C_{\compVec}(\codeVec) = L/\compVec^T \usageVec(\codeVec).
\end{equation*}
Obviously, the capacity region is maximized when there is no coding, i.e., $\codeVec = \minCodeVec\triangleq [k_1;\cdots;k_m]$. We call $C(\minCodeVec)$ the full capacity region.

Similar to our previous discussion for the single-class scenario, we use a M/G/1 queue approximation to model the request queue and use Pollaczek-Khinchin formula to estimate the queueing delay. For a given composition vector $\compVec$, the service time of this M/G/1 queue is modeled by some random variable $X$ whose mean is
\begin{equation*}
\Expect[X] = 1/C_{\compVec}(\codeVec) = \compVec^T \usageVec/L,
\end{equation*}
as per similar reason for non-blocking schemes in single-class scenario. In terms of the second moment, one possibility is to  generalize the Erlang approximation for single-class and consider $X$ to be a mixture of different Erlang random variable: with probability $\alpha_i$, it follows Erlang distribution with parameter $n_i$. While this is doable, it leads to a complicated expression and we believe it will only provide marginal extra insight for the purpose of scheduler design. For this reason, we make a simple and rough assumption that $\Expect[X^2] = \beta \Expect^2[X]$ for some constant $\beta>0$ independent of $\compVec$ and $\codeVec$.
Then the queueing delay is approximated by Pollaczek-Khinchin formula 
\begin{equation*}
\frac{\lambda \Expect[X^2]}{2(1-\lambda \Expect[X])}
= \frac{\beta \lambda \Expect^2[X]}{2(1-\lambda \Expect[X])}
= \frac{\beta \lambda (\compVec^T \usageVec)^2}{2L(L-\lambda \compVec^T\usageVec)},
\end{equation*}
and the expected total queue length (counting all classes) is
\begin{equation*}
Q(\codeVec,\rateVec)=\lambda \frac{\beta \lambda (\compVec^T \usageVec)^2}{2L(L-\lambda \compVec^T\usageVec)} = \frac{\beta ( \rateVec^T \usageVec)^2}{2L(L- \rateVec^T\usageVec)}.
\end{equation*}
We also approximate the service delay of each class $i$  by $D_{s,i}(n_i) = \Delta_i + \sum_{j=0}^{k_i-1}\frac{1}{(n_i-j)\mu_i}$. Noticing that requests of all classes have the same expected queueing delay, we formulate the following optimization problem of finding the best fixed FEC scheme that minimizes the average delay 
\begin{align}
\min_{\codeVec} ~& \sum_{i=1}^m  \alpha_i D_i 
= \frac{\beta \lambda (\compVec^T \usageVec)^2}{2L(L-\lambda \compVec^T\usageVec)}
 + \sum_{i=1}^m \alpha_i D_{s,i}(n_i)  
\label{eq:multi:optimization}\\
\textrm{s.t.} ~& \lambda \compVec^T\usageVec(\codeVec) \le L 
~\textrm{and}
~ n_i \ge k_i-1~\forall~ i=1,\cdots,m \nonumber
\end{align}
Worth pointing out is that we are only interested in the structure of optimal solution and will make use of it for our scheduler design rather than the accurate expression of the solution.
For the following discussion, we will relax the integer requirement for $n_i$'s and allow $n_i$ to be any value $>k_i-1$. 

It is easy to verify that when $\alpha_i>0~\forall i$ the objective of the optimization problem Eq.\ref{eq:multi:optimization} is strictly convex in $\codeVec$.
As a result, we can denote $\codeVec^*(\rateVec)$ as the unique optimal solution for rate vector $\rateVec$. Also let $H(\codeVec) = \{\rateVec|\codeVec = \codeVec^*(\rateVec)\}$. In other words, $H(\codeVec)$ is the union of all rate vectors for which $\codeVec$ is the optimal choice of code lengths. In the case $\codeVec$ is not the optimal for any rate vector, $H(\codeVec)=\{\}$. We say a code vector $\codeVec$ is {\em good} if and only if $H(\codeVec)\neq \{\}$. Theorem \ref{thm:optimal} below is the main result of our analysis.

\begin{theorem}
\label{thm:optimal}
Any good code vector  $\codeVec$ should have the structure
\begin{equation}
\frac{s_i}{\Delta_i\mu_i}= \frac{s_j}{\Delta_j\mu_j}~~ \forall~i,j,
\label{eq:opt:structure}
\end{equation}
where $s_i = \sum_{j=0}^{k_i-1} \frac{1}{(n_i-j)^2}$.
For any such good code vector $\codeVec$, $H(\codeVec)$ is the part of the hyperplane defined by  $\rateVec^T \usageVec(\codeVec) = const(\codeVec)$ within the positive orthant ($\lambda_i>0~\forall i$), where $const(\codeVec)$ is solely determined by $\codeVec$. As a result, while using the optimal  $\codeVec$ at rates $\rateVec\in H(\codeVec)$, the corresponding queue length is a function of only $\codeVec$:
\begin{align*}
Q(\codeVec,\rateVec)|_{\rateVec\in H(\codeVec)} = Q_{opt}(\codeVec)
= \frac{\beta const(\codeVec)^2}{2L(L-const(\codeVec))}.
\end{align*}
\end{theorem}

\begin{IEEEproof} See Appendix.
\end{IEEEproof}

For any pair of good code vectors $\codeVec\neq \codeVec'$, define ordering ``$\prec$'' such that $\codeVec \prec \codeVec'$ if any only if $n_i < n_i'~\forall i$. Similar for ``$\succ$''. 
Also, for two sets of rate vectors $H(\codeVec)$ and $H(\codeVec')$, we say that $H(\codeVec)\prec H(\codeVec')$ if and only if $H(\codeVec)$ is completely contained in the convex hull defined by $H(\codeVec')$ and the origin.

\begin{corollary}
\label{thm:ordered}
The set of all good code vectors is totally ordered with respect to $\prec$.  Moreover, 
 the corresponding rate vector $H(\codeVec)$ and queue length $Q_{opt}(\codeVec)$ are both decreasing functions of $\codeVec$. In other words, 
\begin{equation}
\forall \codeVec \succ \codeVec',~
H(\codeVec)\prec H(\codeVec')~and 
~Q_{opt}(\codeVec)<Q_{opt}(\codeVec').\nonumber
\end{equation}
\end{corollary}

\begin{IEEEproof}
See Appendix.
\end{IEEEproof}

An intuitive interpretation of Theorem \ref{thm:optimal} and Corollary \ref{thm:ordered} is as follows: The full capacity region $C(\minCodeVec)$ is ``sliced'' into layers as hyperplanes $H(\codeVec)$'s. One single (fractional) code vector is optimal for all rates within each layer. When the optimal code vector is used, it produces identical expected queue length throughout the whole layer.  
The layer furthest away from the origin (heavy workload) corresponds to the largest expected queue length. Since the arrival rates are so close to full capacity, any redundancy is detrimental hence no coding should be used. As we move to layers closer to the origin (light workload), the corresponding expected queue length reduces and we can afford to increase the amount of redundancy by using coding.

\begin{figure}[t]
\centering
\includegraphics[width=0.49\columnwidth]{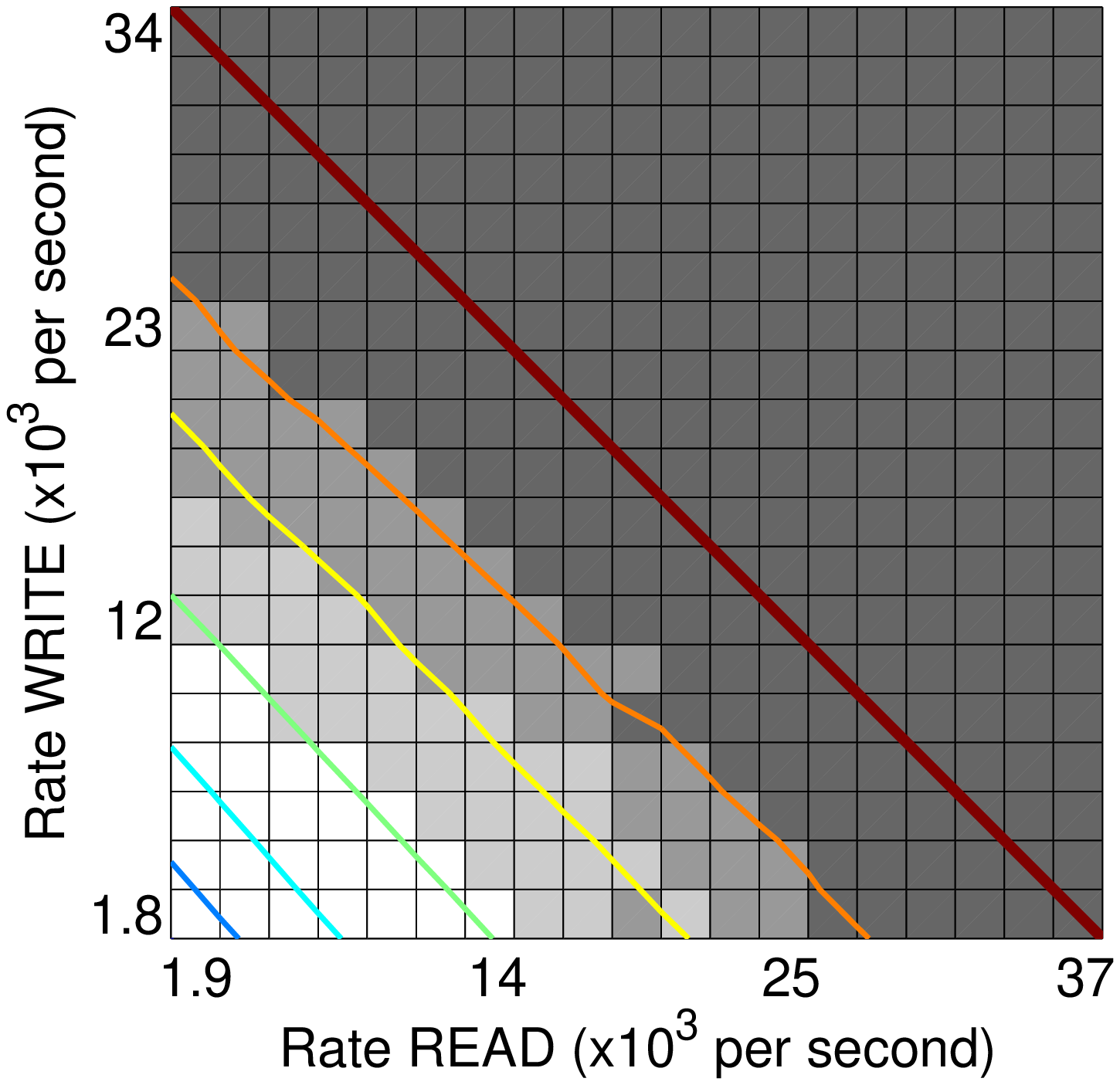}
\includegraphics[width=0.49\columnwidth]{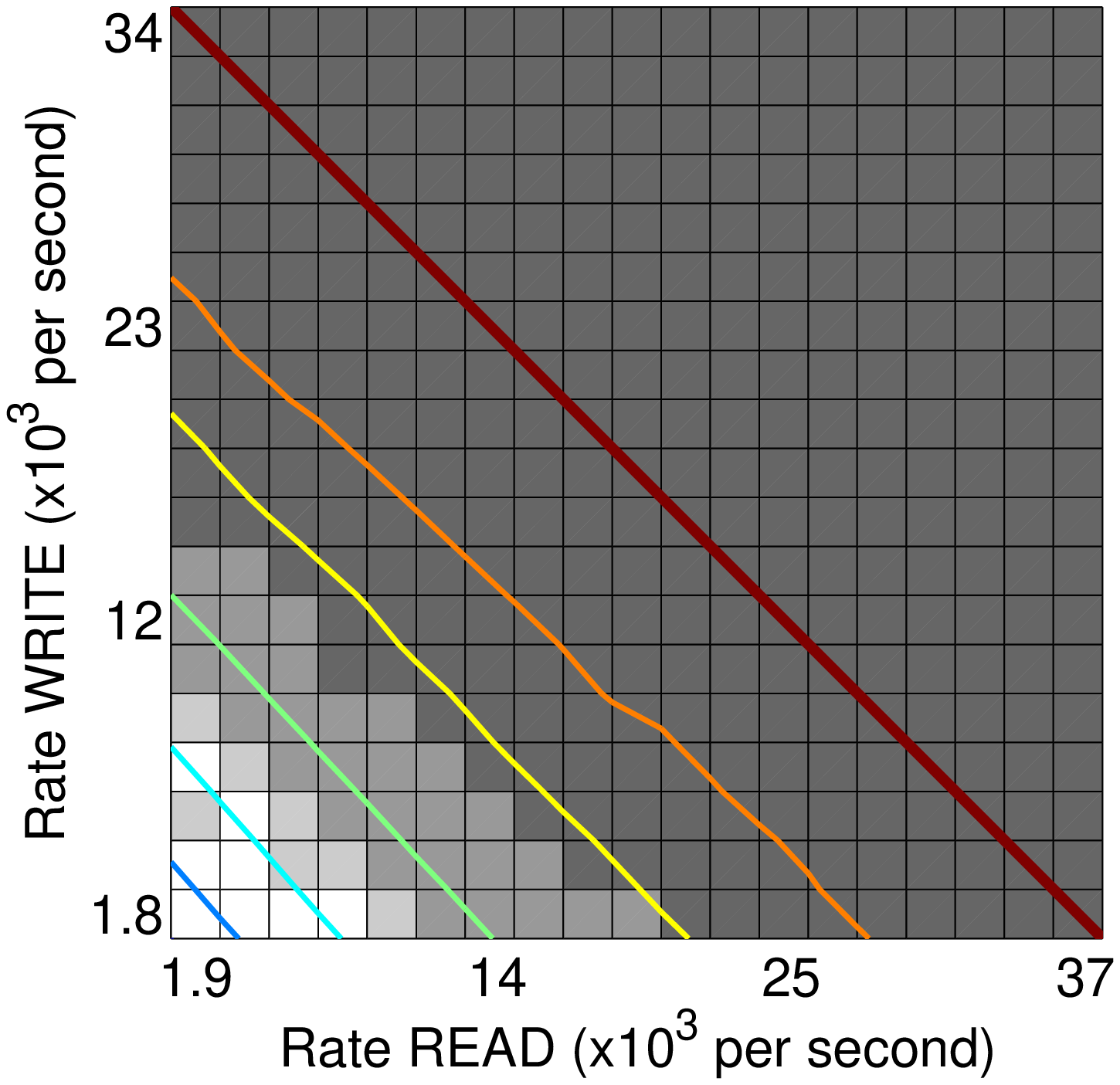}
\vspace{-20pt}
\caption{Best combination of code lengths and queue length, $k_{read}=k_{write}=3$. Left: $n_{read}$. Right: $n_{write}$.}
\label{fig:code_3x3}
\vspace{-5pt}
\end{figure}

\begin{figure}[t]
\centering
\includegraphics[width=0.49\columnwidth]{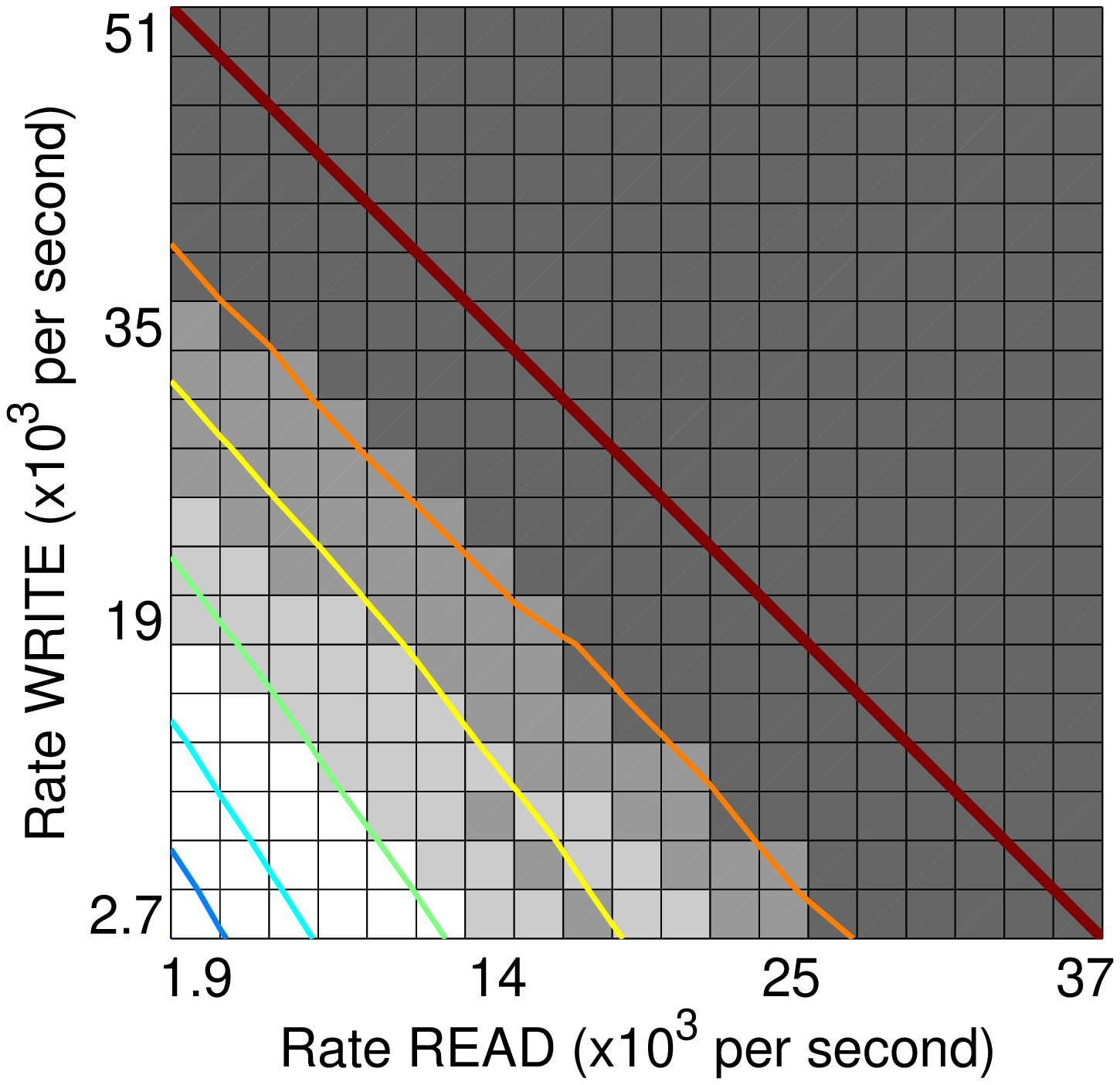}
\includegraphics[width=0.49\columnwidth]{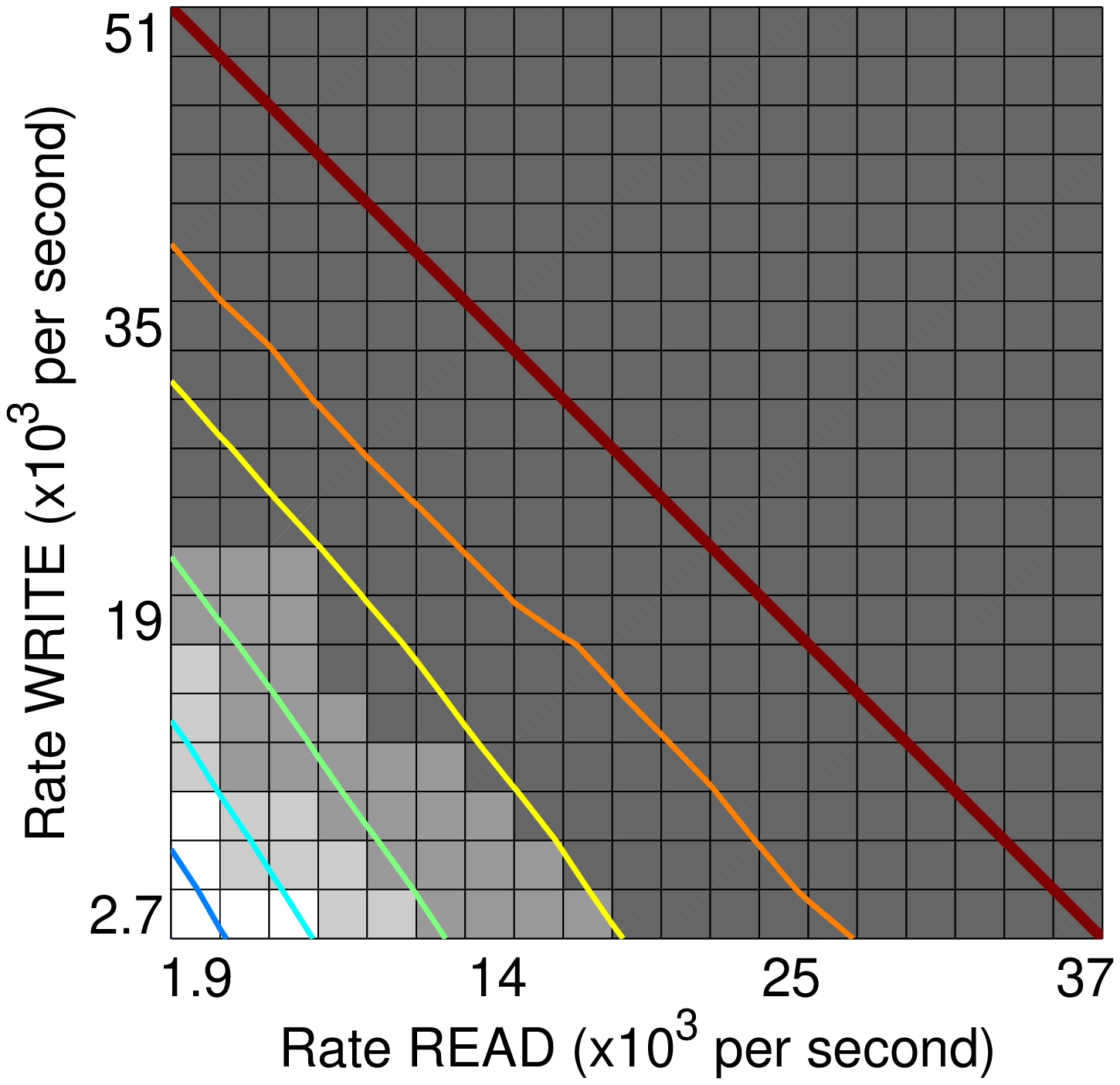}
\vspace{-20pt}
\caption{Best combination of code lengths and queue length, $k_{read}=3$ and $k_{write}=2$. Left: $n_{read}$. Right: $n_{write}$.}
\label{fig:code_3x2}
\vspace{-15pt}
\end{figure}

Remember that Theorem \ref{thm:optimal} and Corollary \ref{thm:ordered} are derived based on non-integer relaxation of code lengths, as well as approximations of the queueing and service delay especially the assumption that $\Expect[X^2] =\beta \Expect^2[X]$. To verify the validity of these results in reality, we perform simulations with $m=2$ classes of requests, literally read and write, with $k_{read}=k_{write}=3$, $n_{read},n_{write}\in\{3,4,5,6\}$ and $L=16$, using traces we collected in March 2012 and chunk size 1MB. We run simulations for at different rate vectors $(\lambda_{read},\lambda_{write})$ with $\lambda_{read}$ and $\lambda_{write}$ varying from $0.05\times$ to $1\times$ of $C_{read}$ and $C_{write}$ respectively, where
$C_{read} = L/k_{read}(\Delta_{read}+1/\mu_{read})$ and 
$C_{write} = L/k_{write}(\Delta_{write}+1/\mu_{write})$ are the maximum arrival rates of read and write request the system can support. At each rate vector, we  run simulations for all $4\times 4$ possible combinations of $(n_{read},n_{write})$, and find the combination that produces the minimum total delay, and record the corresponding average queue length.

 We plot the simulation results in Fig.~\ref{fig:code_3x3}. The x and y axis are the arrival rates of read ($\lambda_{read}$) and write ($\lambda_{write}$)  requests, respectively. The full capacity region is the lower-left half below the diagonal dark red colored line. Beyond this line (top-right) the queue is unstable. Each block in these figures represents one rate vector and the colors of a block represent the combination of $n_{read}$ (left) and $n_{write}$ (right) that results in the smallest total delay among the simulations. Lightest color represents code length of 6 and the darkest represents 3. We also plot contours of queue length levels as colored curves in which points on the same contour/curve have the same average queue length (blue meaning small and orange meaning large).  As we can see, except for a small number of blocks, the rate region is generally divided into 4 layers. Starting from 6 coded blocks in the layer closest to the origin, the number of coded blocks decreases as moving away from the original and eventually becomes 3 in the outmost layer. The small number of blocks of exception near the boundaries are due to the integer constraint on code lengths as well as  randomness in our simulations. Moreover, both the boundaries of these layers and the contours of queue lengths are roughly straight lines, and the boundaries of layers in general are aligned  with the contours of queue lengths at the corresponding arrival rates (some are not shown in the figures). This validates our predictions from Theorem \ref{thm:optimal} and Corollary \ref{thm:ordered} that (1) $H(\codeVec)$ is a hyperplane (which is a line in the 2-dimension space); (2) $Q_{opt}(\codeVec)$ is constant within $H(\codeVec)$; and (3) Both $H(\codeVec)$ and $Q_{opt}(\codeVec)$ are decreasing functions of $\codeVec$. 
Another observation is that as arrival rates increase, $n_{write}$ drops earlier than $n_{read}$ does. This is because, according to our trace, while read and write of 1MB chunks have similar mean task delay (both around 140 ms), $\Delta_{write}$ is much larger than $\Delta_{read}$ (114 ms vs. 61 ms), and as we discuss before in Section \ref{sec:singletype}, the queueing delay starts to dominate at lower utilization with larger $\Delta$. 

It appears in Fig.~\ref{fig:code_3x3} that all contours of queue length are roughly parallel to the boundary of the full capacity region (the diagonal dark red line), which may suggest the illusion of $H(\codeVec)$ being parallel to full the capacity boundary. We would  point out that this is just a coincidence. In Fig.~\ref{fig:code_3x2} we plot the results for $k_{read}=3$, $k_{write}=2$, and $n_{write}\in\{2,3,4,5\}$. It is clear in this case that the contours are not parallel to the full capacity boundary, especially for low arrival rates.

\subsection{\multiclass -- Multi-class Backlog-based Adaptive FEC}
\label{ssec:multi:algorithm}

An important implication of Corollary \ref{thm:ordered} is that there is a one to one mapping from $Q_{opt}$ to the corresponding good code vector $\codeVec$, since the set of good code vectors is totally ordered and $Q_{opt}$ is an strictly decreasing function of the good code vectors. Roughly speaking, the larger $Q_{opt}$ is, the smaller (good) code vector should be. This suggests that generalizing the single-class scheduler \nonBlocking to accommodate multiple classes of requests is plausible. 

A natural and intuitive way of generalizing \nonBlocking is to first enumerate the set of good code vectors using the structure of good code vectors provided by Eq.\ref{eq:opt:structure}, then sort these code vectors and solve for the corresponding backlog thresholds for every pair of consecutive code vectors as we did for the single-class scheduler \nonBlocking. At last, depending on which range between the thresholds the backlog size falls into, we pick the corresponding $\codeVec$.  However, this approach is not quite feasible when the number of classes $m$ is large, mainly due to the integer requirement for $\codeVec$. Notice that Eq.\ref{eq:opt:structure} can be converted into a polynomial equation of $n_i$ and $n_j$, each of degree $2k_i$ and $2k_j$ respectively. A straightforward way of finding the set of good codes is to first pick the code length for a certain class, say $n_1$ without loss of generality, to be an integer under consideration, then solve Eq.\ref{eq:opt:structure} numerically for the corresponding code lengths of the other classes. However, the solutions obtained by doing this are not necessarily integers. In fact, they will most likely be non-integers unless the values of $\Delta_i,\mu_i,k_i$'s happen to pair up perfectly. So for every such fractional solution of $\codeVec$ (except for $n_1$), we need to decide which of $\lfloor n_i \rfloor$ and $\lceil n_i \rceil$ to pick, for all $i\neq 1$. There is no obvious way to solve this other than enumerating all $2^{m-1}$ potential solutions, computing the expected delays and picking the best one. So the computational complexity is exponential in $m$ for each integer value of $n_1$. Such exponential complexity may be affordable for static algorithms which assume statistics of task delays ($\Delta$ and $\mu$) to be fixed. But in reality delay statistics of cloud storage systems vary over time and need to be updated regularly in order to harvest the best performance. More importantly, stale delay statistics can be dangerous because if they are too optimistic compared to reality then the scheduling algorithm will tend to allocate more tasks per request than it  should, which will result in large backlog and queueing delay. In such cases, the exponential complexity is forbiddingly expensive. 

In fact, the exponential complexity of computing the backlog thresholds can be avoided. The key is to observe that $Q_{opt}$ is also a decreasing function of each individual $n_i$ and there is also a one to one mapping from $Q_{opt}$ to  $n_i$, assuming the other classes are using the corresponding optimal code lengths. So instead of adapting $\codeVec$ as a whole, adaptation can be done for each $n_i$ separately. So instead of computing one set of $\sum_{i=1}^m(n_i^{max}-k_i)$ backlog thresholds across which a transition in the code vector $\codeVec$ occurs, we compute one smaller set of $n_i^{max}-k_i$ thresholds for each class $i$ individually across which a transition in only $n_i$ occurs. Here $n_i^{max}$ denotes the maximum number of tasks allowed for a class-$i$ request. These two approaches should produce the same set of thresholds but the separated approach avoids the combinatorial problem of enumerating the set of good code vectors at the first place. Denoting $\{Q_{i,k_i},\cdots,Q_{i,n_i^{max}}\}$ as the set of thresholds computed for class $i$, the pseudo-code for the \multiclass scheduler we develop using the separate approach is as follows:

\vspace{4pt}
\hrule
\vspace{2pt}
\noindent \textit{\textbf{\multiclass (Multi-class Backlog-based Adaptive FEC)}}
\hrule
\vspace{2pt}

\noindent  Do the following for every request $r$

\begin{algorithmic} [1]

\STATE $\overline{Q} \leftarrow $ backlog size upon arrival of request $r$.
\STATE $i\leftarrow$ class that the  request $r$ belongs to.
\STATE Find $n$ such that $\overline{Q}\in [0,Q_{i,n})$ for $n=n_i^{max}$, or $\overline{Q}\in [Q_{i,n}, Q_{i,n-1})$, or $\overline{Q}\in [Q_{i,n},\infty)$ for $n=k_i$.
\STATE Serve request $i$ with an $(n,k_i)$ code when it becomes HoL.
\end{algorithmic}
\hrule

~

To compute the set of thresholds for each class, recall that $Q_{opt}(\codeVec)$ stays fixed for all $\rateVec\in H(\codeVec)$ according to Theorem \ref{thm:optimal}. So it suffices to consider rate vectors along a certain direction specified by a fix composition vector $\compVec$ and find the crossover backlog sizes along that direction. In particular, for class $i$, we consider the direction along the $i$-th axis. In other words, we consider the class-$i$-only arrival case with $\alpha_i = 1$ and $\alpha_j=0~\forall j\neq i$. In the example of Fig.~\ref{fig:code_3x3}, this is equivalent to finding the 
intersections for the boundaries of layers with the x axis (read-only arrival) in the left plot for the thresholds of read requests, and finding the  intersections with the y axis (write-only arrival) for write requests. Further, noticing that \multiclass behaves identically to \nonBlocking when arrivals are single-class, these intersections with the $i$-th axis can be computed using Eq.\ref{eq:threshold}, with parameters $\Delta_i,\mu_i,k_i$, just as we do  for \nonBlocking in the previous section.

\begin{figure*}[!ht]
\centering
	\subfigure[Read Heavy -- Average Delay]{
		\label{fig:9:ave}
		\includegraphics[width=\smallwidth]{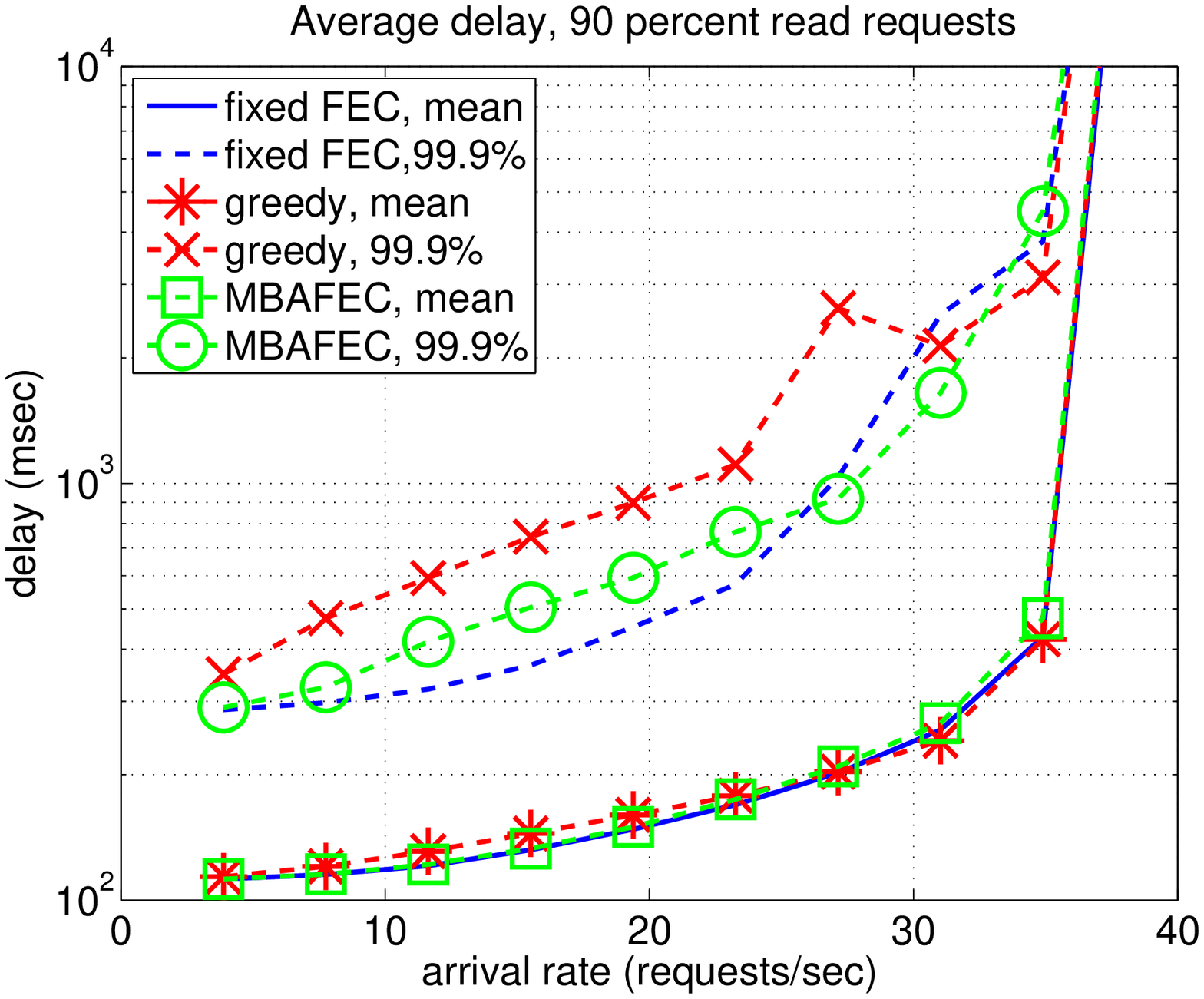}
	}%
~~~
	\subfigure[Read Heavy -- Read Delay (normalized)]{
		\label{fig:9:read}
		\includegraphics[width=\smallwidth]{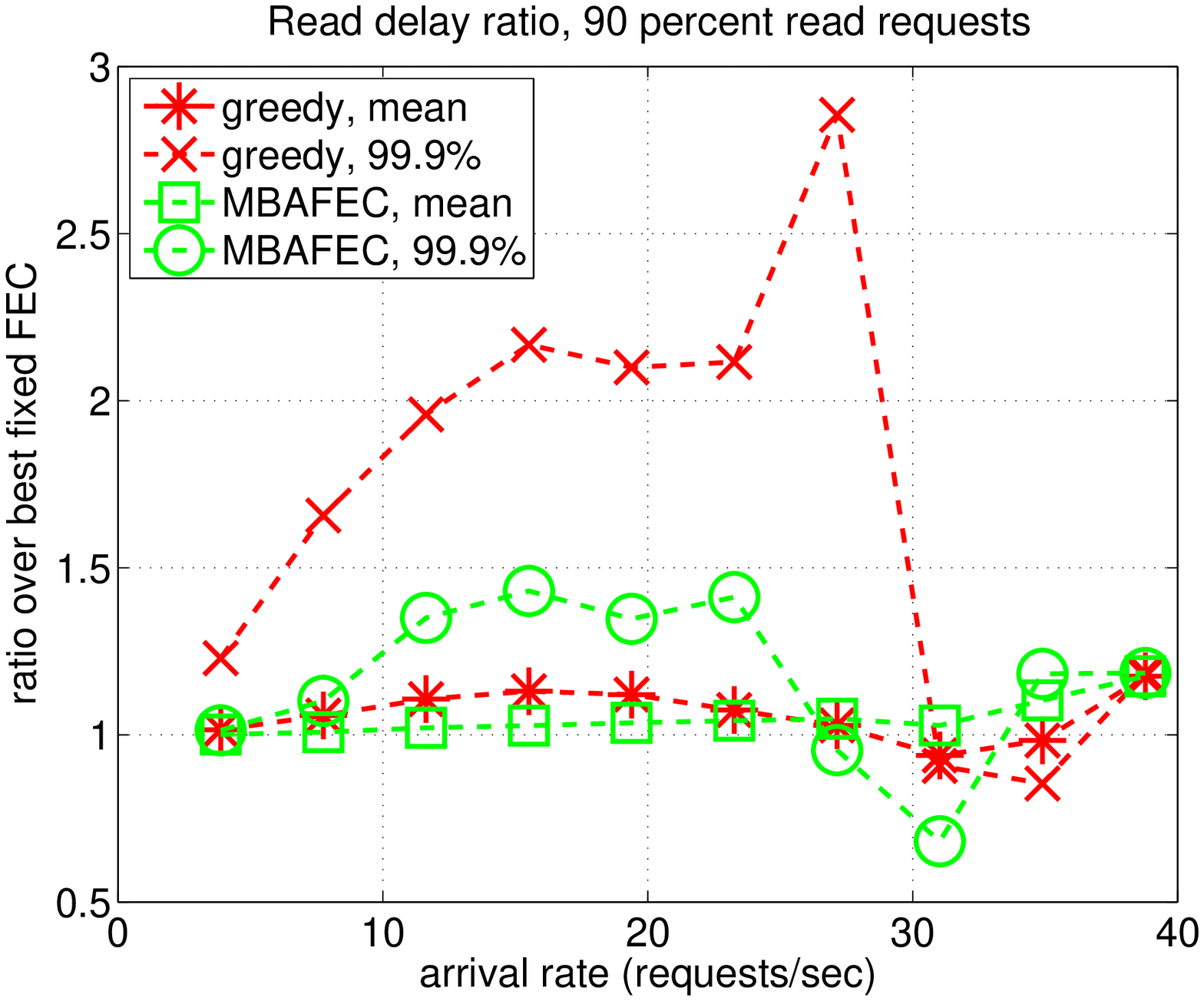}
	}
~~~
	\subfigure[Read Heavy -- Write Delay (normalized)]{
		\label{fig:9:write}
		\includegraphics[width=\smallwidth]{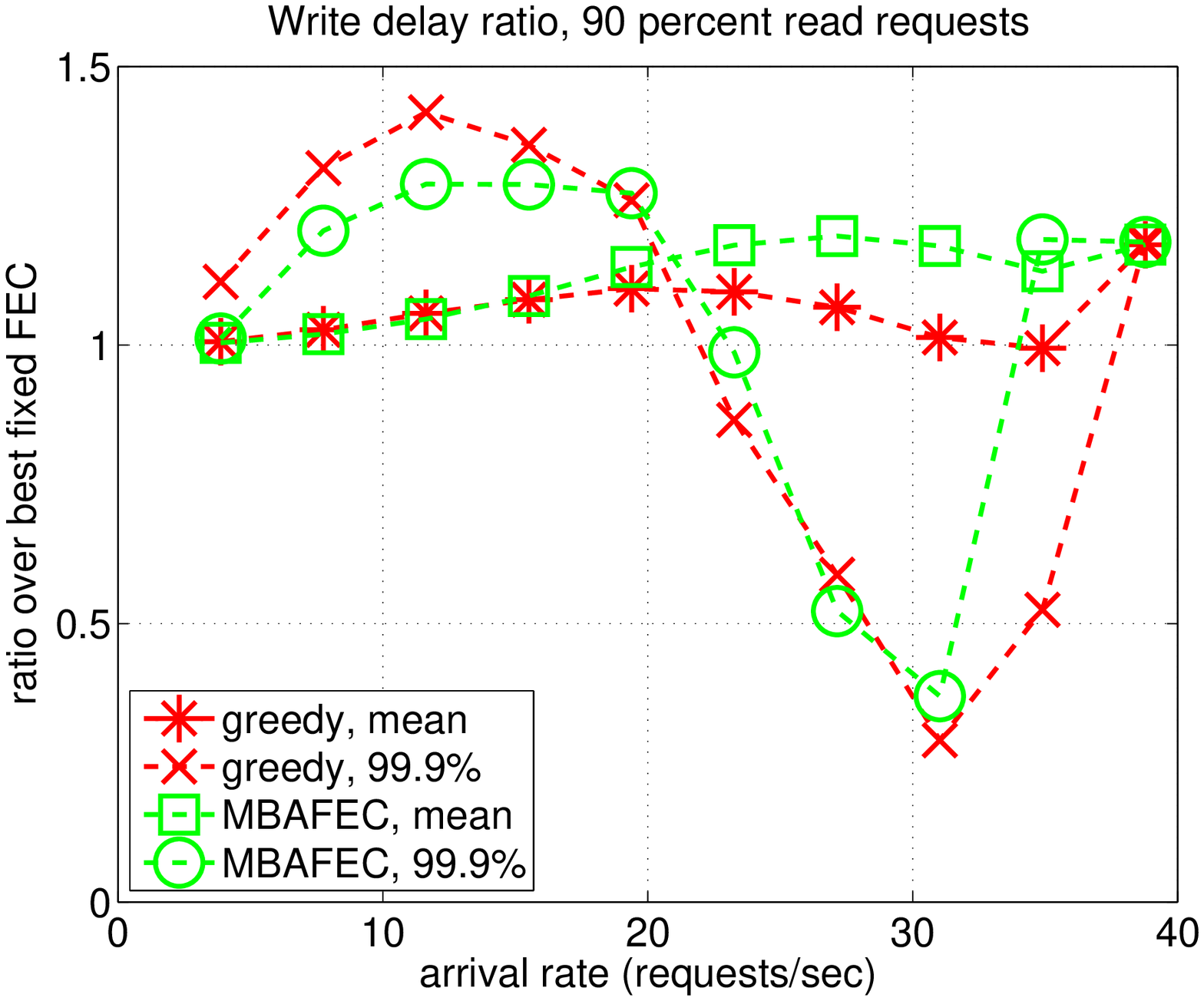}
	}
\\
\subfigure[Balanced -- Average Delay]{
		\label{fig:5:ave}
		\includegraphics[width=\smallwidth]{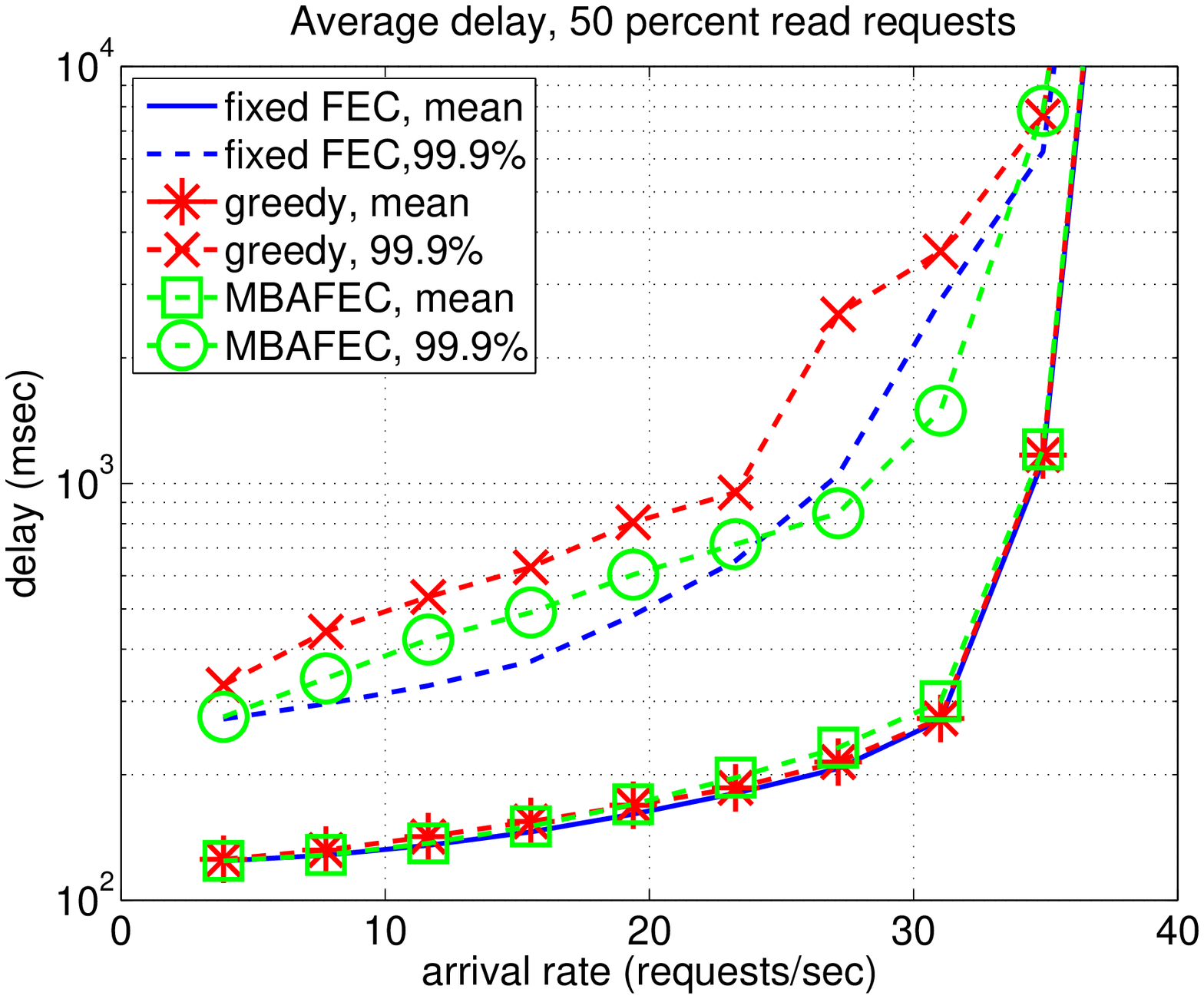}
	}%
~~~
	\subfigure[Balanced -- Read Delay (normalized)]{
		\label{fig:5:read}
		\includegraphics[width=\smallwidth]{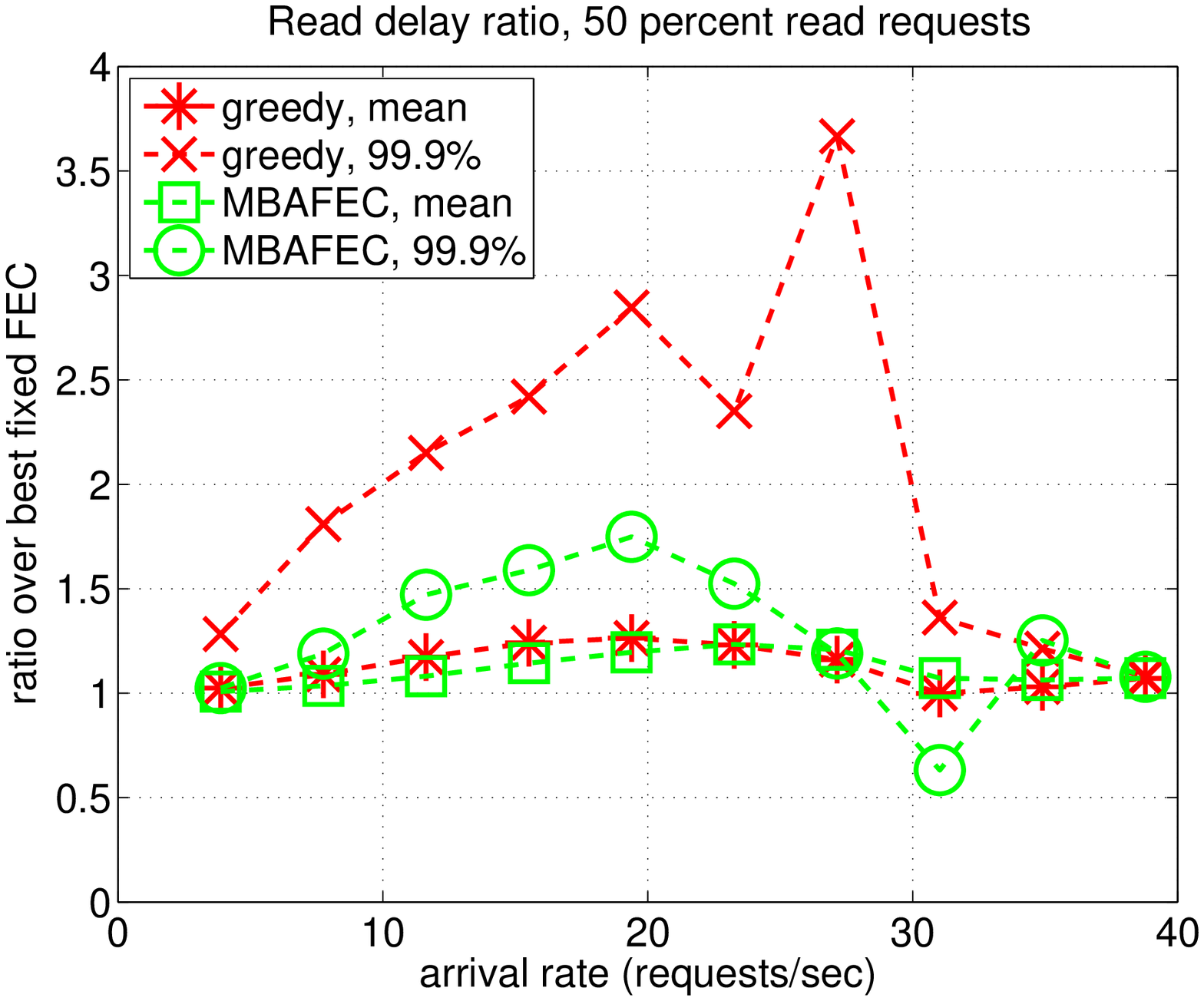}
	}
~~~
	\subfigure[Balanced -- Write Delay (normalized)]{
		\label{fig:5:write}
		\includegraphics[width=\smallwidth]{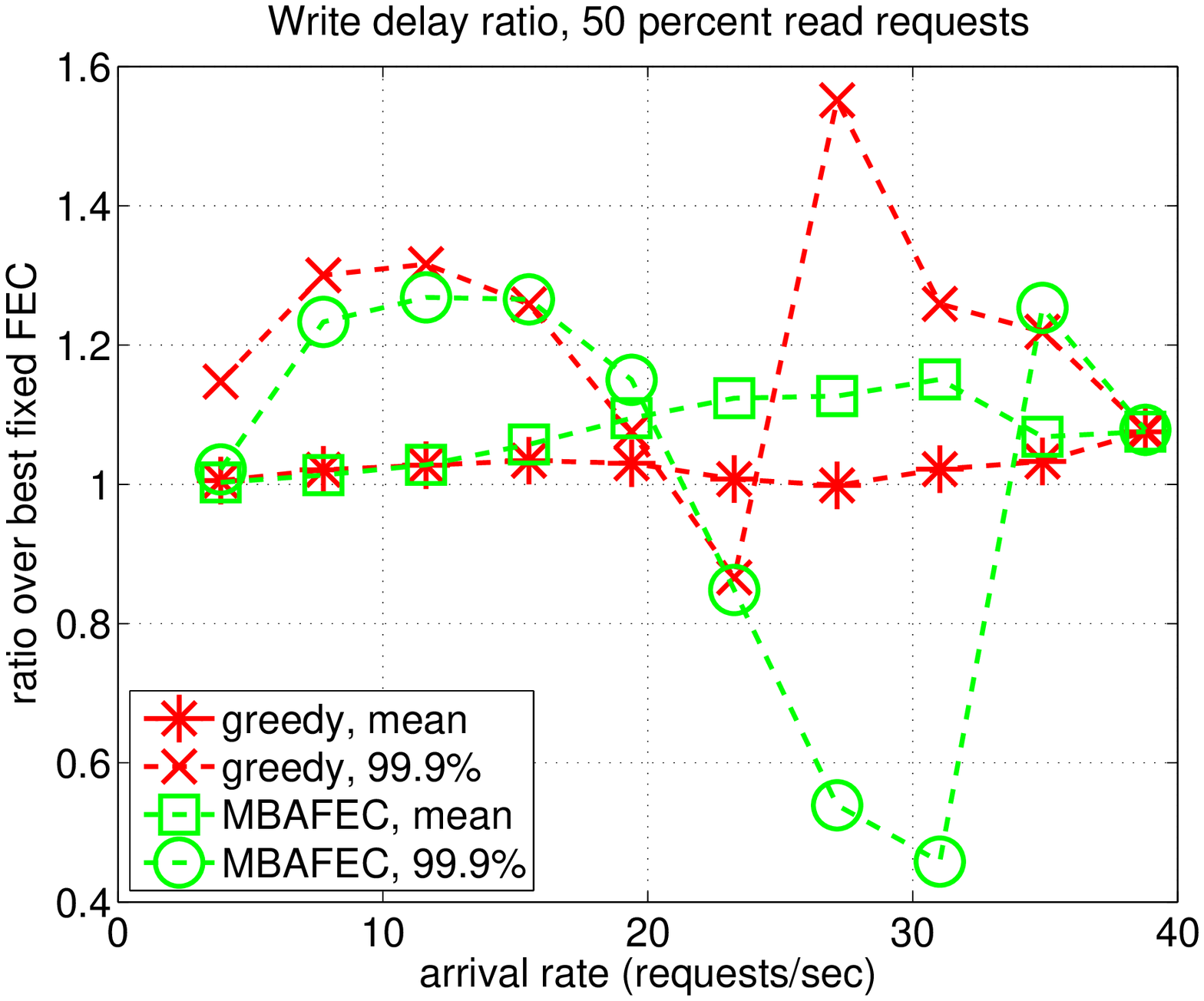}
	}
\\
~~~
\subfigure[Write Heavy -- Average Delay]{
		\label{fig:1:ave}
		\includegraphics[width=\smallwidth]{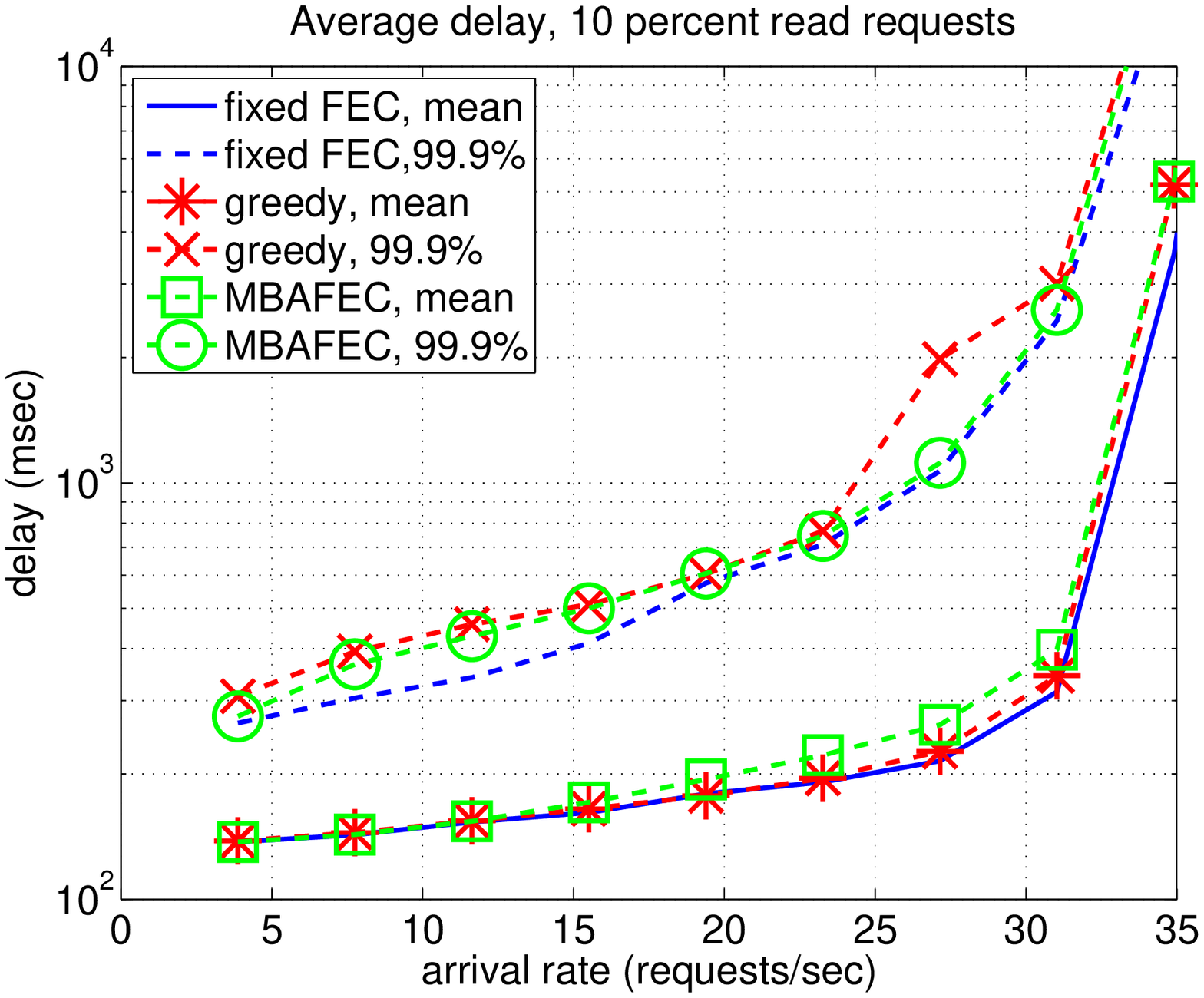}
	}%
~~~
	\subfigure[Write Heavy -- Read Delay (normalized)]{
		\label{fig:1:read}
		\includegraphics[width=\smallwidth]{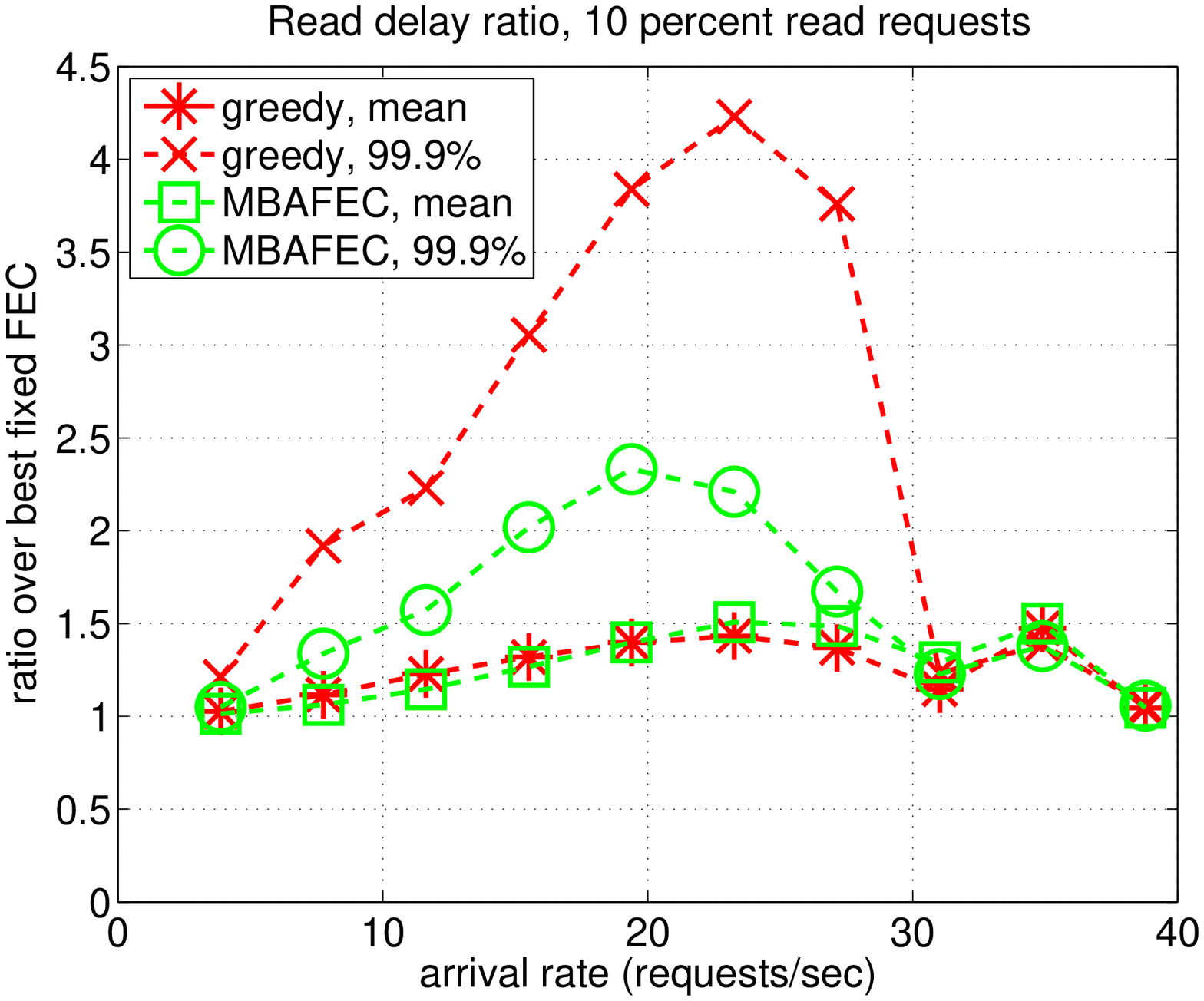}
	}
~~~	\subfigure[Write Heavy -- Write Delay (normalized)]{
		\label{fig:1:write}
		\includegraphics[width=\smallwidth]{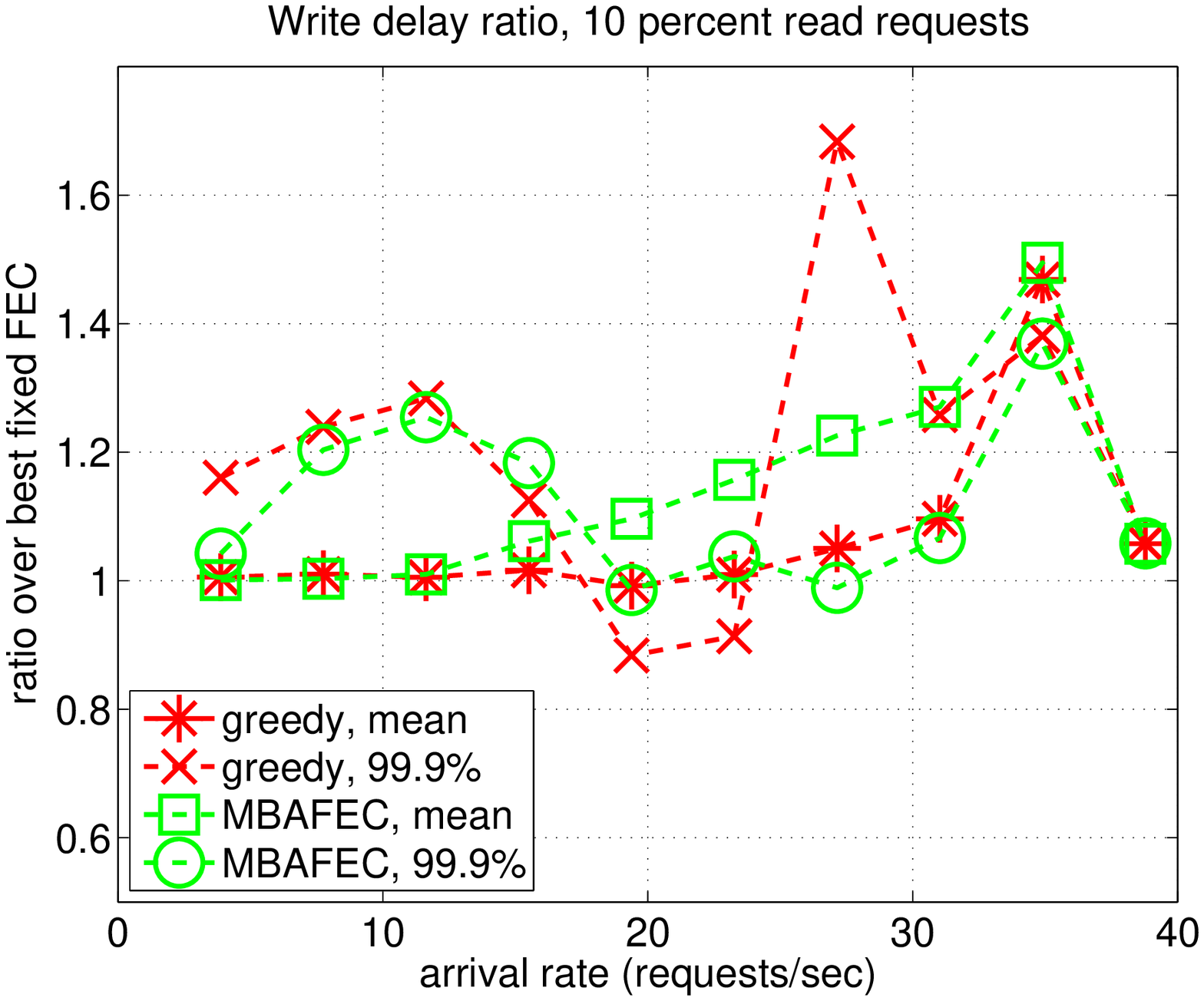}
	}
\vspace{-10pt}
\caption{Delay performance from simulations}
\label{fig:multi:delay}
\vspace{-15pt}
\end{figure*}

\subsection{Performance Evaluation of \multiclass}
For performance evaluation, we perform simulations with $m=2$ classes of requests, literally read and write, with $k_{read}=k_{write}=3$, $n_{read},n_{write}\in\{3,4,5,6\}$ and $L=16$, using traces we collected in March 2012 and chunk size 1MB. We simulated three scenarios: read heavy ($\alpha_{read} = 0.9$), balanced ($\alpha_{read} = 0.5$), and write heavy ($\alpha_{read}=0.1$). We also extended Greedy to accommodate multiple classes: each class-$i$ request uses $(\min(l,n_i^{max}),k_i)$ or $(k_i,k_i)$ code, depending on the number of idle threads $l$ upon arrival.

Fig.~\ref{fig:multi:delay} illustrates the delay performance for \multiclass and Greedy. We also run simulations with fixed FEC scheme with all 16 combinations of code lengths at every arrival rate of each scenario and use the best average delays ($\alpha_{read}D_{read,Y} + \alpha_{write}D_{write,Y}$ with $Y$ = mean  and 99.9\%. $D_{read,Y}$ and $D_{write,Y}$ represent the mean and 99.9\% delay for read and write requests), the best delays (mean and 99.9\%) for read requests, and the best delay (mean and 99.9\%) for write requests as baselines. We want to point out here that the combinations of code lengths that result in the best average delay, read delay, and write delay are not necessarily the same. We observe in our simulations that the combination that results in best read delay usually uses a large code length for read requests and the minimum code length for write requests $n_{write} = k_{write}$, which results in high write delay. It is the opposite observation for codes that produce the best write delay. The combination that produces the best average delay is usually in between.
So in these figures, we are comparing 6 delay metrics of one adaptive scheme \multiclass (or Greedy) against multiple fixed FEC schemes, each of which excels in one particular delay metric.

In the left column of Fig.~\ref{fig:multi:delay} we plot the average delays. Similar to the results for the single class case, both \multiclass and Greedy perform well and achieves roughly the same average mean delays as the best fixed FEC schemes throughout the full capacity region. \multiclass also achieves the lower envelop of fixed FEC schemes in terms of  average 99.9\% delay and outperforms Greedy. More interesting are the middle and right columns, in which we plot the read and write delays of \multiclass and Greedy, normalized by the best corresponding delays with fixed FEC. \multiclass and Greedy perform similarly in terms of mean delays and both stay within $1.5\times$ of the best mean delays with fixed FEC. Remember this comparison is made against the fixed FEC scheme that produces the best mean read or write delay, which is different from the one that produces the best average delay.
These two adaptive schemes perform quite differently in terms of 99.9\% delays. For 99.9\% write delay, \multiclass and Greedy are similar and stay within $1.5\times$ of the best fixed FEC for most arrival rates in all three scenarios. On the other hand, \multiclass constantly outperforms Greedy significantly in terms of 99.9\% read delays in all three scenarios. \multiclass stays within $1.5\times$, $1.8\times$ and $2.4\times$ of the best delay from fixed FEC in read heavy,  balanced and write heavy scenarios respectively, while Greedy can perform as bad as $2.9\times$, $3.7\times$ and $4.2\times$ in each scenario.   
There are two reasons for such difference of performance in read and write requests. Firstly, in our trace read operations have a much larger delay spread than write operations have. As a result, read requests benefit significantly by reducing service delay from parallelism with appropriately chosen code length, while write requests cannot benefit much due to its smaller delay spread. More importantly, Greedy is ``class-oblivious'' and it does not make use of the difference in delay statistics of different classes of requests in deciding the code length for each class. 

\begin{figure}[t]
\centering
\includegraphics[width = 0.7\columnwidth]{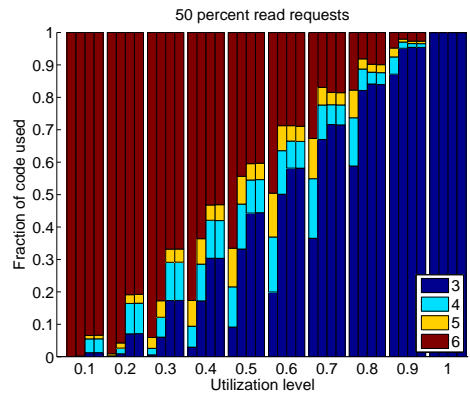}
\vspace{-10pt}
\caption{Composition of code lengths. Each group is ordered as [\multiclass read, \multiclass write, greedy read, greedy write].}
\label{fig:code_compo}
\vspace{-15pt}
\end{figure}

To better understand how \multiclass and Greedy behave differently, we plot the code composition (the fractions of requests served by different code lengths) of read and write requests using \multiclass and Greedy from 10\% to 100\% utilization levels. Fig.~\ref{fig:code_compo} shows the code compositions for the balanced arrival scenario (plots for read/write heavy scenarios are similar). At each utilization level, the four bars represent the code compositions of read requests with \multiclass, write requests with \multiclass, read requests with Greedy, and write requests with Greedy, from left to right. For each bar, the colors represent the fraction of requests served with code length 3, 4, 5 and 6, from bottom to top. Generally speaking, both schemes behave as expected: at low utilization, both schemes mostly use code length 6 since service delay dominates;  as utilization increases, both become less aggressive and increase the fraction of requests served by smaller code lengths; at very high utilization, both reduce to no coding for both read and write requests ($n_{read}=k_{read}$ and $n_{write}=k_{write}$). The major difference we observe between \multiclass and Greedy is that the code compositions for read and write requests differs significantly with \multiclass except for at very low and very high utilization levels, while they are almost identical with Greedy at all utilization levels. Remember that in Greedy, the code length used to serve a request is determined by the number of idle threads upon arrival of the request and the range of code lengths allowed to serve the request. Since we assume Poisson arrivals, both read and write requests should statistically observe the same distribution of number of idle threads. Also because both read and write requests have the same range of code lengths in our simulations, they result in having the same code composition. If different types of requests have different ranges of code lengths, the code compositions will be slightly different for the edge cases (not enough idle threads or too many idle threads). On the other hand, \multiclass treats read and write requests very differently, given that read and write operations have very different delay distributions. For read requests, since delay of read operations has a small fixed component ($\Delta_{read}$) and a large exponential tail ($\mu_{read}$), the overhead in queueing delay of parallelism is much smaller than the benefit from service delay. So \multiclass is more aggressive in using large code lengths ($n_{read} > 3$). For write requests, since write operations has a large fixed delay component, \multiclass is more conservative. For medium to high utilization levels, \multiclass is even more conservative than Greedy for write requests (\multiclass serves fewer write requests with $n_{write}\ge 5$ than Greedy does at 80\% to 100\% utilization). 
We also observe that at all utilization level, Greedy serves most  requests with either the maximum or minimum value of $n$ while \multiclass serves a much larger fraction of requests with medium values of $n$. This all-or-nothing behavior of Greedy is the main reason for its poor performance at high percentile delays, since the service delay distribution of simple chunking without coding ($n=k>1$) is only slightly better than doing nothing ($n=k=1$).

\section{Conclusion and Future Directions}
\label{sec:conclusion}
We presented novel solutions that combine parallel thread scheduling and FEC for accessing data stored in public clouds substantially faster in the sense of mean, 90th percentile, 99th and higher percentile latencies. The solutions can be applied to other distributed data storage technologies that exhibit high delay variations for object or block storage. 

In the analysis of the problem, we admitted a mixed traffic load with multiple classes of files read/write requests. But, chunk and file sizes of each class were predetermined and fixed. 
In general, better performance might be achieved if chunk size is also adaptable. For example, smaller $k$ could extend the capacity region at high utilization, and larger $k$ may reduce service delay at low utilization. Extending the adaptation schemes in this paper to incorporate adaptive chunk sizing is the next step in our research plan.

In our work, we neglected the dollar amount cost of using redundant requests, e.g., Amazon S3 charges 0.01\$ per 1000 requests for PUT, COPY, POST, or LIST Requests and 0.01\$ per 10,000 requests for GET and all other requests. For now, by limiting the code rate and level of chunking, we put upper bounds on these costs in our work. Since not all parts of data are delay sensitive, such costs can be managed by applying our techniques on a smaller fraction of the load (e.g., initial segments of a video file). Extensions to capture the cloud pricing in the problem formulation and devise scheduling schemes accordingly are part of our ongoing work.

\appendix[Proofs of Theorem \ref{thm:optimal} and Corollary \ref{thm:ordered}]

\begin{IEEEproof} 
First observe that the first term of the objective approaches $\infty$ as $\lambda\compVec^T\usageVec\rightarrow L$ and the second term approaches $\infty$ as $n_i\rightarrow k_i-1$. Since both terms are lower bounded by 0, it follows that the objective approaches $\infty$ at the boundary of the feasible region. Together with the fact that the objective is a strictly convex function of $\codeVec$, it follows that for any given feasible $\rateVec$,  the optimal solution $\codeVec^*(\rateVec)$ is strictly inside the feasible region. Since the objective is differentiable, its partial derivative equals $0$ only at $\codeVec^*$. In other words, if for some $\codeVec$ 
\begin{align*}
\frac{\partial D}{\partial n_i} &= \frac{\partial}{\partial \compVec^T \usageVec}\left(\frac{\beta \lambda (\compVec^T \usageVec)^2}{2L(L-\lambda \compVec^T \usageVec)}\right) \frac{\partial \compVec^T \usageVec}{\partial n_i} +  \alpha_i\frac{\partial D_{s,i}}{\partial n_i}\\
&= \frac{\beta \alpha_i\Delta_i}{2L}\left(\frac{L^2}{(L-\lambda \compVec^T \usageVec)^2} - 1\right) - \frac{\alpha_i}{\mu_i}s_i
\end{align*}
equals to 0 for all $i$, then $\codeVec = \codeVec^*(\rateVec)$. 
Here $s_i = - \mu_i \frac{\partial D_{s,i}}{\partial n_i} = \sum_{j=0}^{k_i-1} \frac{1}{(n_i-j)^2} $. This condition is equivalent to 
\begin{equation}
\frac{L^2}{(L-\lambda \compVec^T \usageVec)^2} - 1= \frac{2L}{\beta}\frac{s_i}{\Delta_i\mu_i}~~\forall i.
\label{eq:opt:condition}
\end{equation}
Due to the uniqueness of the optimal solution, the other direction is also true: for any given good code vector $\codeVec$, if $\rateVec$ satisfies Eq.\ref{eq:opt:condition}, then $\codeVec = \codeVec^*(\rateVec)$ or equivalently $\rateVec\in H(\codeVec)$. 

An important property of good code vectors implied by Eq.\ref{eq:opt:condition} is that all good code vectors line up on the curve specified by 
\begin{equation}
\frac{s_i}{\Delta_i\mu_i} = \frac{s_j}{\Delta_j\mu_j}~~ \forall~i,j.
\label{app:eq:opt:structure}
\end{equation}
Given this, for any good $\codeVec$, denote $\pi(\codeVec) = \frac{2L}{\beta}\frac{s_i}{\Delta_i\mu_i}$ for any $i$, when Eq.\ref{app:eq:opt:structure} is satisfied. Then Eq.\ref{eq:opt:condition} can be rewritten as
\begin{equation}
\rateVec^T\usageVec = L - L/\sqrt{1+\pi(\codeVec)}\triangleq const(\codeVec).
\end{equation}
In other words,
$H(\codeVec) = \{\rateVec|\rateVec^T\usageVec(\codeVec) = const(\codeVec)\}.$

It is obvious that $\pi(\codeVec)$ is strictly decreasing of $n_i>k_i-1$, for all $i$. So $\pi()$ is invertible and for any $a > b$ in the range of $\pi()$ we have
$
\pi^{-1}(a)\prec \pi^{-1}(b).
$
This implies that the good code vectors are totally ordered in  decreasing order of $\pi()$. 

Consider any two good code vectors $\codeVec \succ \codeVec'$. For any $\rateVec\in H(\codeVec)$, $\rateVec^T\usageVec(\codeVec) = const(\codeVec)$. 
Note that $const(\codeVec)$ is a strictly increasing function of $\pi(\codeVec)$, so it is a strictly decreasing function of $\codeVec$. Hence $const(\codeVec)<const(\codeVec')$, and we have
$
\rateVec^T\usageVec(\codeVec')  < \rateVec^T\usageVec(\codeVec)
= const(\codeVec)
< const(\codeVec').
$
The first inequality is due to the fact that both $\rateVec$ and $\usageVec$ are $>0$. Now we can conclude that any $\rateVec\in H(\codeVec)$ is strictly within the convex hull defined by $H(\codeVec')$ and the origin. So $H(\codeVec)\prec H(\codeVec')$.

It is easy to verify that $Q_{opt}(\codeVec)$ is an increasing function of $const(\codeVec)$. Since $const(\codeVec)$ is a decreasing function of $\codeVec$, $Q_{opt}(\codeVec)$ is a decreasing function of $\codeVec$. 
\end{IEEEproof}

\bibliographystyle{IEEEtran}

\begin{IEEEbiography}[{\includegraphics[width=1in,height=1.25in,clip,keepaspectratio]{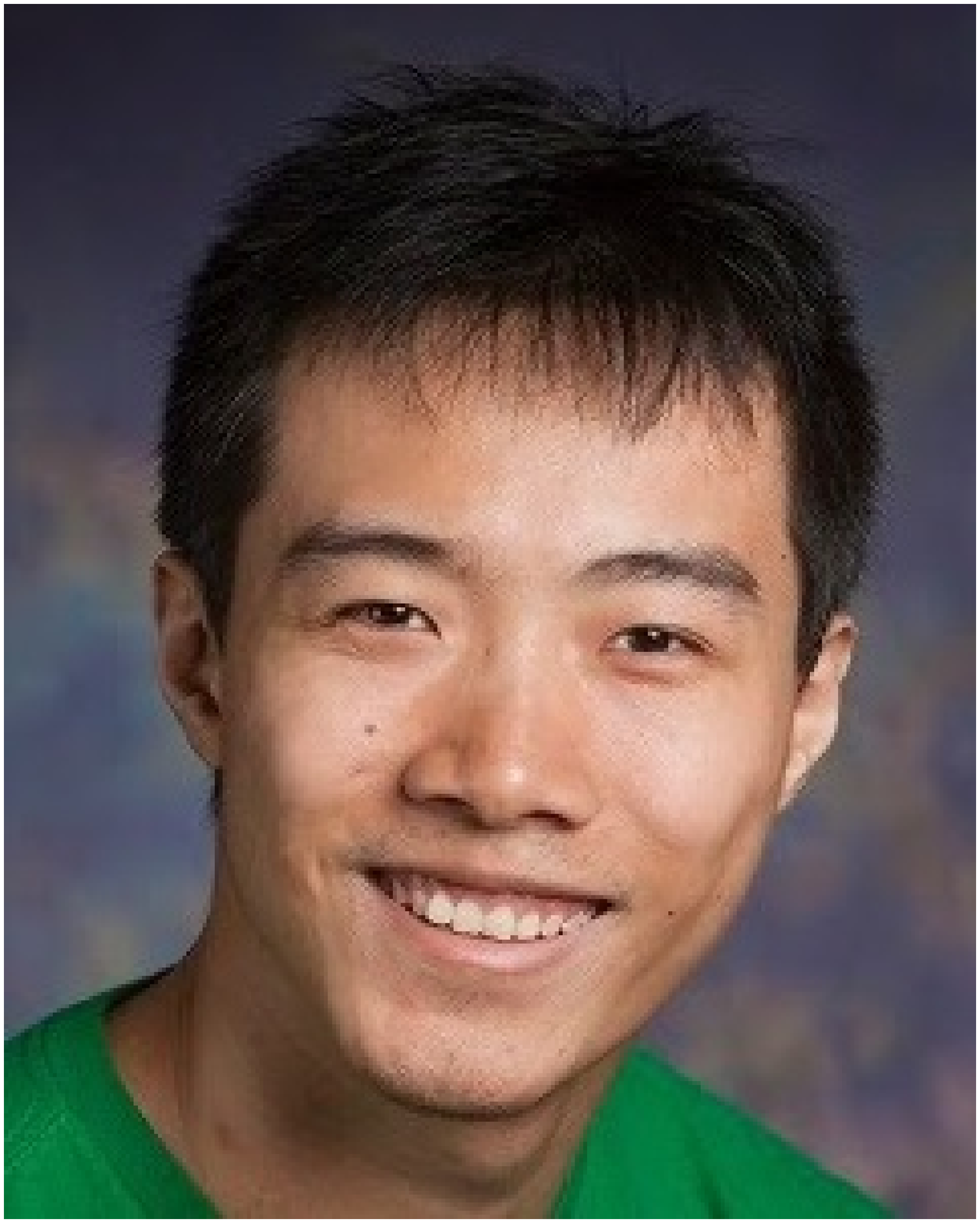}}]{Guanfeng Liang}
(S'06-M'12)
received his B.E. degree from University of Science and Technology of China, Hefei, Anhui, China, in 2004, M.A.Sc. degree in Electrical and Computer Engineering from  University of Toronto, Canada, in 2007, and Ph.D. degree in Electrical and Computer Engineering from the University of Illinois at Urbana-Chanpaign, in 2012. 
He currently works with DOCOMO Innovations (formerly DOCOMO USA Labs), Palo Alto, CA, as a Research Engineer. 
\end{IEEEbiography}

\begin{IEEEbiography}[{\includegraphics[width=1in,height=1.25in,clip,keepaspectratio]{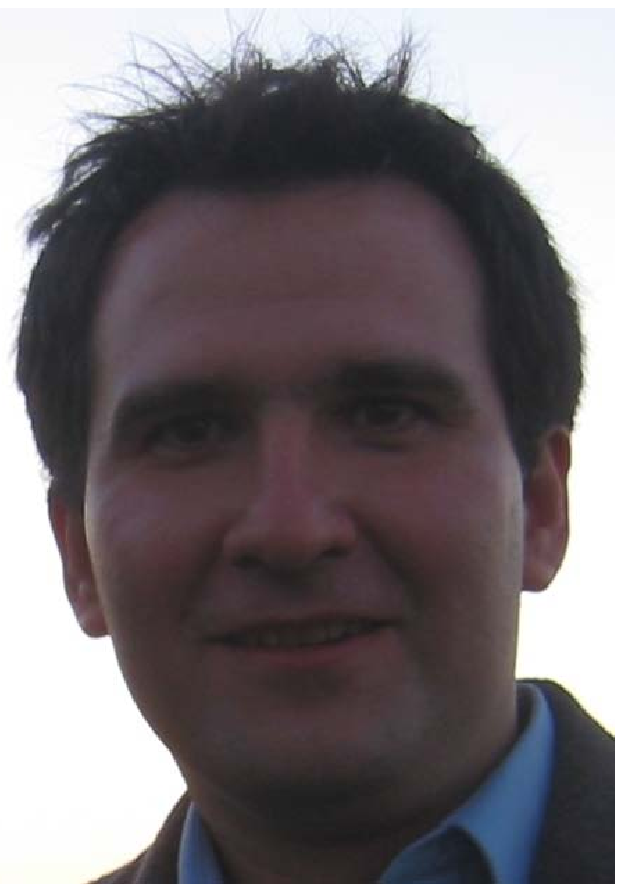}}]{Ula\c{s}~C.~Kozat} 
(S’97-M’04-SM’10) received his
B.Sc. degree in Electrical and Electronics Engineering
from Bilkent University, Ankara, Turkey, in 1997,
M.Sc. degree in Electrical Engineering from the
George Washington University, Washington, DC, in
1999, and Ph.D. degree in Electrical and Computer
Engineering from the University of Maryland,
College Park, in 2004. He currently works at DOCOMO Innovations (formerly DOCOMO USA Labs), Palo Alto, CA, as
a Principal Researcher.

\end{IEEEbiography}
\end{document}